\documentclass{aa}  

\usepackage{graphicx}
\usepackage{placeins}
\usepackage{natbib}
\usepackage{txfonts}
\usepackage{ulem}
\usepackage{pdflscape}
\usepackage{amsmath}
\usepackage[colorlinks=true,citecolor=blue]{hyperref}
\makeatletter

\makeatletter
\renewcommand*\aa@pageof{, page \thepage{} of \pageref*{LastPage}}
\makeatother

\newcommand{\MJup}{M$_{\mathrm{Jup}}$\xspace}

\newcommand{\MSun}{M$_{\odot}$\xspace}

\newcommand{\kms}{km\, s$^{-1}$\xspace}

\begin{document}

   \title{A study of the frequency and characteristics of stellar companions and Jupiter-like planets in nearby open clusters }


   \author{R. Gratton\inst{1}, M. Bonavita\inst{1,2}, D. Mesa\inst{1}, S. Desidera\inst{1}, A. Zurlo\inst{3,4,5}, S. Marino\inst{6}, V. D'Orazi\inst{1,7}, E. Rigliaco\inst{1}, V. Nascimbeni\inst{1}, D. Barbato\inst{1}, G. Columba\inst{1,8}, V.  Squicciarini\inst{1,9} 
   }
  \institute{
   \inst{1}INAF-Osservatorio Astronomico di Padova, Vicolo dell'Osservatorio 5, Padova, Italy, 35122-I \\
   \inst{2}Institute for Astronomy, University of Edinburgh Royal Observatory, Blackford Hill, EH9 3HJ, Edinburgh, UK \\
   \inst{3}Instituto de Estudios Astrof\'isicos, Facultad de Ingenier\'ia y Ciencias, Universidad Diego Portales, Av. Ej\'ercito 441, Santiago, Chile\\
   \inst{4}Escuela de Ingenier\'ia Industrial, Facultad de Ingenier\'ia y Ciencias, Universidad Diego Portales, Av. Ej\'ercito 441, Santiago, Chile\\
   \inst{5}Millennium Nucleus on Young Exoplanets and their Moons (YEMS)\\
   \inst{6}Department of Physics and Astronomy, University of Exeter, Stocker Road, Exeter, EX4 4QL, UK\\
   \inst{7}Dipartimento di Fisica, Universit\`{a} di Roma Tor Vergata, via della Ricerca Scientifica 1, 00133, Roma, Italy\\
   \inst{8}Dipartimento di Fisica e Astronomia G. Galilei, Universit\`{a} di Padova, Via Francesco Marzolo, 8, 35121 Padova, Italy\\
   \inst{9}LESIA, Observatoire de Paris, Universit\'e PSL, CNRS, Sorbonne Universit\'e, Universit\'e Paris Cit\'e, 5 place Jules Janssen, 92195 Meudon, France
} 

   \date{Received; accepted }

 
  \abstract
   { Observations of companions of solar-type stars in nearby young moving groups (NYMGs) show that they split into two groups: stellar and brown dwarf companions (mass ratio $q>0.05$) and Jupiter-like (JL) planets ($q<0.02$). The frequency of JL planets in NYMGs appears to be higher than that obtained from radial velocity (RV) surveys.}
   { We extended the search for companions to three nearby clusters of intermediate age: Hyades, Coma Berenices, and Ursa Major. They are older and formed in more massive events than the NYMGs.}
   { The sample of host stars is complete for the core of the clusters, while we considered only a fraction of the tidal tails. We used the same methods considered for the members of NYMGs.}
   { We obtained a fairly complete sample of stellar companions and detected six massive JL planets. We found a lower frequency of equal-mass companions than in the NYMGs; this might be related to how binaries form in these environments. We also observed a concentration of stellar binaries in the cores of Ursa Major and Coma Berenices; we attribute this to the selective loss of low-mass systems. The observed scarcity of wide companions in Hyades can be due to the destruction of binaries in close encounters. The frequency of JL planets is lower than in the NYMGs but similar to that obtained from RV surveys. This extends the correlation with age and mass previously found for NYMGs.}
   { Results of this study alone do not indicate whether age or mass are the factors driving the observed correlation. A comparison of the frequencies of free-floating planets from microlenses and in young associations favours mass as the main driving parameter. Once the initial cluster mass function is considered, the frequency of JL planets in NYMGs is consistent with the results obtained using RVs.}


\titlerunning{Companions around intermediate-age stars }
\authorrunning{R. Gratton et al.}

   \maketitle
%

\section{Introduction}

Recent progress in high-contrast imaging (HCI), high-precision radial velocities (RVs), space astrometry, and photometry are providing a wealth of data on companions to nearby solar-type stars. When combined with information on the membership to clusters and young associations, these data provide a fairly complete view of the population of companions down to the planetary regime and over a very wide range of separations for stars with relatively secure ages and birthplaces. These unprecedented databases provide a very useful comparison sample for models of the formation of companions. This is particularly useful when considering young populations because they are less affected by the dynamical evolution of the systems. In previous studies \citep{Gratton2023, Gratton2023b, Gratton2024}, we used this sample to explore the formation of companions to stars in very nearby young moving groups (NYMGs) over a relatively wide range of masses. The main result of these works is that the companion's mass ratio $q$ distribution has two main peaks: one at a high value ($q>0.05$) that typically includes stellar secondaries and a second one at a low value ($q<0.02$) that typically includes planetary companions. Stellar secondaries are distributed almost uniformly over a wide range of separations, while planetary companions typically have a semi-major axis $a$ that is $a<100$ au. Since the mass ratio and separation of this group of planetary companions are similar to those of Jupiter, we refer to this group as Jupiter-like (JL) planets; in our discussion, they have $0.001<q<0.02$ and $1.5<a<50$ au. We can consider the gap between the two groups as the extension at larger separations of the well-known brown dwarf (BD) desert found in RV surveys at shorter separations \citep{Marcy2000, Raghavan2010, Stevenson2023, Unger2023}. This reinforces the idea that massive BDs are the low end of the mass distribution of stellar companions and the low-mass ones can be considered as the high-mass tail of the planetary companions. The existence of two well-separated populations of companions to stars strongly suggests that they formed through different mechanisms: The stellar companions likely formed from core fragmentation at separations larger than 500 au \citep{Offner2010, Offner2016} and by disk fragmentation at shorter separations \citep{Kratter2010}, while the JL planets may be explained by the formation of gas giant planets through core or pebble accretion \citep{Pollack1996, Mordasini2012, Bitsch2015}.

When focusing on the JL planet population in different NYMGs, \citet{Gratton2023b} found a surprisingly high frequency of JL planets in the $\beta$ Pic moving group. \citet{Gratton2024} found a trend of decreasing frequency of this class of objects with increasing age and or mass of the association when considering other NYMGs. Given the existing correlation between the presence of giant planets on wide orbits and smaller inner planets (e.g. \citealt{Rosenthal2022}), this datum can be related to the site of formation of the Solar System. We can attribute the trend observed by \citet{Gratton2024} to a more difficult formation of JL planets in hostile environments characterised by the presence of massive stars \citep{Winter2022} and more frequent close encounters with other stars \citep{Adams2001, Adams2004, Wang2021, Winter2018, Parker2021, ConchaRamirez2021, Wright2022}. These effects may produce trends with the mass of the star-forming region. Alternatively, the observation by \citet{Gratton2024} can be due to the progressive destruction of planetary systems with time \citep{Cai2017,Flammini2019, Li2019,Li2020b,Li2020, Maraboli2023}, that is, to a long-term dynamical effect. This last mechanism is more likely to occur for closely packed multi-planetary systems (which are likely to be quite common: \citealt{Fang2013, Weiss2018, Humphrey2020, Obertas2023, Turtelboom2024}) where even small orbital changes due to external perturbations can cause ejection of some or even all of the planets \citep{Gotberg2016, Petit2020, Zurlo2022}. These mechanisms may explain why we obtained a frequency of JL planets in NYMGs that is much higher than that found in RV surveys, which target stars that are typically old and likely formed in much wider groups than the nearby NYMGs (e.g. \citealt{Lada2003}). In fact, we obtained a frequency of JL planets in the range between 20\% and more than 100\% in NYMGs \citep{Gratton2024}, but we estimated a frequency of $5.8\pm 2.2$\%  by extrapolating the recent results by \citet{Fernandes2019} to the range of mass and semi-major axis of our interest (see Appendix~\ref{appendix:rv_frequency}). Although the correction for incompleteness bears considerable uncertainty, the discrepancy between the results obtained for NYMGs and those from RV surveys is significant and needs explanation.

Since environment and dynamical evolution are likely to play an important role in the formation and evolution of companions to solar-type stars, extending this study to older clusters and associations in the solar neighbourhood is the obvious continuation of the work done in \citet{Gratton2024}. Open clusters are likely to be more representative of the typical environment in which most stars form. For this reason, the results presented in this paper can be considered as a bridge between the result obtained for NYMGs and RVs. Of course, one should keep in mind that the low-density associations considered in this series of papers are likely to disperse over hundreds of parsec on a time scale of the order of 100 Myr due to the effect of the Milky Way gravitational field \citep{Roeser2019, Jerabkova2021}, and it is therefore very difficult to separate their members from neighbouring field stars. In addition, there are no clusters younger than 400 Myr and a total mass larger than a few tens of \MSun within a 100 pc distance to the Sun (see \citealt{Cantat-Gaudin2018,Moranta2022})\footnote{Hereafter, we call solar neighbourhood the region within 100 pc from the Sun.}, where we may consider the search for low-mass companions quite complete, and the sample size may be large enough for a statistical discussion. The closest open cluster younger than this age limit is in fact the Pleiades at 136 pc \footnote{\citet{Ortega2007} proposed a possible connection between the Pleiades and the NYMG AB Dor; however, the recent analysis by \citet{Heyl2022} based on the data from the Gaia satellite found very few objects associated with the Pleiades within a distance of 100 pc of the Sun.}. Self-gravity is generally important for the identified groups older than 300 Myr in the solar neighbourhood (Hyades = Melotte 25, Coma Berenices = Melotte 111, and Ursa Major = Collinder 285); the cores are considered as open clusters in the two early cases. The Ursa Major moving group should have been an open cluster in the past and the core is still clearly recognisable \citep{Soderblom1993}. These systems underwent a strong dynamical evolution after their formation, causing important selection effects on their companion population. 

Even with such strong caveats, a study of the distribution of companions to stars belonging to these three groups remains useful. In fact, most stars in the solar neighbourhood are believed to have formed in similar environments \citep{Lada2003, Winter2022b}, which were likely characterised by a higher density and total mass than NYMGs. Since most open clusters dissolve into the general field on a time scale between $10^6$ to $10^8$ yr (e.g. \citealt{Lamers2005}), a large fraction of the (typically old) field stars observed by RV surveys might indeed have originated in open clusters. In fact, if we compare the current star formation rate in clusters in the solar neighbourhood of $520\pm 70$ \MSun kpc$^{-2}$ Mpc$^{-1}$ considered by \citet{Lamers2005} with the overall current star formation rate of $790\pm 160$ \MSun kpc$^{-2}$ Mpc$^{-1}$ by \citet{Bonatto2011}, we derive that $66^{+18}_{-14}$\% of the stars in the solar neighbourhood formed in open clusters. A slightly lower value of 40\% was obtained by \citet{Piskunov2008}.

The paper is structured as follows: In Section~\ref{sec:group_selection} we discuss the main properties of the stellar groups considered in the study. The detection of stellar and substellar companions among members of the core of the groups and their tails within about 100 pc from the Sun is presented in Section~\ref{sec:companion_detection}. Section~\ref{sec:companion_parameters} presents the derivation of the main parameters (masses and orbital semi-major axis) of the individual components. Section \ref{sec:substellar_companions} discusses the substellar companions found among the programme stars. We discuss the completeness of the search in Section~\ref{sec:completeness}.  Section~\ref{sec:discussion_stellar} contains the discussion of the results obtained for the stellar companions and Section \ref {sec:discussion_jl} contains the same for JL planets. Conclusions are drawn in Section~\ref{sec:conclusions}.

\section{Group selection and their properties}
\label{sec:group_selection}

We selected for this study the three intermediate-age stellar groups (Hyades, Coma Berenices, and Ursa Major) that have a large population of members within 100 pc from the Sun\footnote{Very recently, \citet{Gagne2023} recognised the existence of the 500 Myr old Oceanus nearby moving group. However, there are only six solar-type stars that are known to be members of this group; only three of them are single stars. The statistical significance of the results that concern this small group is low. We then prefer not to use it in this study.}. Although they are in different evolutionary stages and two of them may no longer be gravitationally bound, for simplicity we refer to them as intermediate-age open clusters. They have ages in the range $415-625$ Myr. The distance limit was chosen to allow reasonable completeness in the search for companions over a wide range of $q$ and $a$ from astrometry and HCI since both methods are very sensitive to distance. As for the NYMGs, we only considered stars with masses $>0.8$ \MSun. The main characteristics of these open clusters are listed in Table \ref{tab:associations}. 

\subsection{Cluster age and member selection}

The typical time scale for dynamical evolution in clusters is given by the relaxation time, which for Hyades is currently estimated to be 35 Myr \citep{Roeser2011}, and should be even less for the two other clusters because they have much fewer members. As the relaxation time is much shorter than the age, which is estimated at 625 Myr \citep{Perryman1998} for Hyades, the clusters here considered are highly dynamically evolved. For this reason, we expect that a large fraction of the stars in these clusters were lost since their formation. For these open clusters, we then considered both core members as well as (a fraction of) those objects that were originally associated with the cluster but now are lost in the field. In particular, we also included members of the Latishev 2 and Mecayotl 1 moving groups that may be associated with Coma Berenices (Melotte 111) though they may be slightly younger (about 400 Myr rather than about 600 Myr: \citealt{Olivares2023}). In this respect, we note that a precise age estimate of $533\pm 42$ Myr for Coma Berenices was recently obtained using the contact binary 12 Com \citep{Lam2023}, reducing this difference. This estimate agrees within the errors with the estimate of $562\pm 83$ Myr by \citet{Silaj2014}. The age of the Ursa Major cluster (Collinder 285) has been estimated to be $414\pm 23$ Myr \citep{Capistrant2024}, in good agreement with previous estimates \citep{King2003}.

We considered members of each cluster or association from the sources listed in Table \ref{tab:associations}. As in \citet{Gratton2024}, we kept only those systems for which the primary stars have a mass $>0.8$ \MSun. We checked all stars for membership in the respective clusters using the BANYAN $\Sigma$ code \citep{Gagne2018}\footnote{\url{https://www.exoplanetes.umontreal.ca/baNYMGn/banyansigma.php}}. The list of stars of Coma Berenices was already carefully checked for membership by \citet{Olivares2023}. In addition, membership determined by the BANYAN $\Sigma$ code is not really meaningful for the Ursa Major association, where only core members verify the membership criteria considered in BANYAN $\Sigma$. Although the Ursa Major group space velocity is distinct from those of other young field stars \citep{Soderblom1990} and there should be relatively few interlopers to contaminate the sample, the membership of several of the members considered here may still be questioned (e.g. the case of Sirius: \citealt{Soderblom1993, King2003}). However, in our opinion, this does not affect the statistical meaning of the results obtained in this paper as long as the age of the star is close to the age of the Ursa Major group. The list of members of Hyades includes 169 stars. 144 of these stars have a membership probability $>0.5$ according to Banyan $\Sigma$. For ten of the remaining 25 stars, the low membership probability is due to the fact that Banyan $\Sigma$ uses a radial velocity (RV) that is very different from that of the cluster. Most of them are short-period spectroscopic binaries, and the RV used by Banyan $\Sigma$ is questionable. If we replace the RV considered by Banyan with that given by Gaia or neglect the RV, the membership probability for these objects would be high. The 15 stars that anyhow have low membership probability according to Banyan $\Sigma$ have proper motion that, however, is much more similar to that of Hyades than the value typical for field stars and lie within 12 pc from the cluster centre.

\subsection{Distances}

In this paper we considered the European Space Agency astrometric Gaia satellite $G$ \citep{Gaia_DR3} and Two Micron All Sky Survey 2MASS $K$ magnitudes \citep{2mass}. Whenever possible, we adopted distances from the Gaia data release 3 (DR3); otherwise, from Gaia Data Release 2 (DR2) or, as a last resort, from Hipparcos \citep{vanLeeuwen2007}. Interstellar absorption was neglected, considering the close distance of the target stars. 

The distance values listed in Table \ref{tab:associations} are the median of the values of the individual core members with mass $>0.8$ \MSun obtained from the parallaxes listed in Gaia DR3 \citet{Gaia_DR3}. The Gaia limit reported in this table is the mass of a BD with $G=19$ at the mean distance and age of each association. We obtained this using the isochrones from \citet{Baraffe2015} for the age of the associations\footnote{ \citet{Baraffe2015} computes isochrones for masses $>0.040$ \MSun for ages in the 400-625 Myr range, which is considered in this study. They do not cover the entire brown dwarf range but are appropriate for the purpose described in this section.}. It provides a first estimate of the completeness in the search for wide stellar and BD companions using Gaia. Given the moderately large age of the clusters and associations considered here, these limits are quite close to the dividing mass between BDs and low-mass stars.

\begin{table*}
\caption{Main parameters for intermediate-age clusters and associations.}
\scriptsize
\begin{tabular}{lcccccccccl}
\hline
Group      & Other &Distance & Age  & [Fe/H] & $N_{\rm Prim}$ & $N_{\rm Prim}$ (core)& $M_{\rm Core}$ & $M_{\rm Orig}$ & Gaia$_{\rm lim}$ & Source     \\
           & & (pc)    & (Myr)& &$>0.8$\MSun &$>0.8$\MSun &\MSun&\MSun&\MSun&            \\
\hline
Ursa Major & Collinder 285 & 26.3   & 415  & 0.00 &   64 &  13 & 55 &  $415\pm 165$ & 0.058 &\citet{King2003,Gagne2018b} \\
           &&        & &     &     &  &     &      &       &   \citet{Nakajima2010, Elsanhoury2020}\\
Coma Berenices       & Melotte 111 & 90.6   & 530  & 0.00 &  119 &42 &  87 &  $587\pm 245$ & 0.083 &\citet{Olivares2023}\\
Hyades     & Melotte 25 & 46.5   & 625  & 0.13 &  169 & 129 & 450 & $1200\pm 400$ & 0.078 &\citet{Perryman1998,Lodieu2019}\\
\hline
\normalsize
\end{tabular}
\tablefoot{$N_{Prim}$ is the number of primaries with a mass $>0.8$ \MSun considered in this paper; $M_{\rm Core}$ is the present mass of the core of the cluster; $M_{\rm Orig}$ is the original mass of the cluster derived as described in the text; and Gaia$_{\rm lim}$ is the mass of a star or BD in the cluster corresponding to a magnitude of $G=19$.}
\label{tab:associations}
\end{table*}

\subsection{Masses}
\label{sec:masses}

The masses for the main-sequence stars are derived from the absolute Gaia $G$ magnitudes using the calibration by \citet{Pecaut2013} if $M_G<3$, else from the isochrone by \citet{Baraffe2015} of appropriate age (these isochrones are only available for stars less massive than 1.4 \MSun). In case the star is an unresolved binary in the Gaia Catalogue, we corrected the Gaia $G$ magnitudes for the contribution by the secondary before extracting the masses. We did this in an iterative way, but convergence was always fast. The masses for red giants (that is, stars that are not in the main sequence and are then not covered by the calibration by \citealt{Pecaut2013}) were obtained from the PARSEC isochrones \citep{Costa2019a, Costa2019b, Nguyen2022} \footnote{\url{http://stev.oapd.inaf.it/PARSEC/papers.html}} considering ages that are appropriate for each cluster or association.

The sum of the masses of the components considered in this study (primary mass $>0.8$ \MSun) for Hyades cluster (all objects within 30 pc from the centre of the cluster) is 319 \MSun with a total of 210 \MSun being in the primaries and the rest in the other components of each individual system\footnote{This total includes a total mass of 23 \MSun that is the contribution of the progenitors of the eight Hyades white dwarfs listed by  \citet{Salaris2018,Pasquini2023}. A single white dwarf is known in Coma Berenices \citep{Dobbie2009}, while one (Sirius B) is among the Ursa Major members considered in this study.}. The same totals are 196 \MSun and 154 \MSun for the sum of Coma Berenices, Latishev 2 and Mecayotl 1; 161 \MSun and 120 \MSun for Ursa Major (including association members). The current masses of the clusters are larger because we should also consider the contribution by lower-mass stars. We used the mass function of Hyades within 30 pc of the centre by \citet{Roeser2011} to estimate that the observed mass should be multiplied by a factor of about 1.5 to consider this effect. This would lead to a total mass of 478 \MSun for Hyades. This value is slightly higher than that of 435 \MSun obtained by \citet{Roeser2011} within 30 pc of the cluster centre. This difference may be due to a different consideration of the contribution by the companions. 

If we use the same correction factor for low-mass stars obtained for Hyades, the total mass for the combination of Coma Berenices, Latishev 2 and Mecayotl 1 is 294 \MSun. This value agrees fairly well with the values given by \citet{Olivares2023} who estimated masses of $87\pm 4$ \MSun for the core, $74\pm 4$ \MSun for the tail, $92\pm 5$ \MSun for Latyshev 2, $75\pm 5$ \MSun for Mecayotl 1. The sum of these estimates gives a total of 328 \MSun. Finally, adopting the same correction factor of 1.5 for the Ursa Major group (that is, stars identified through BANYAN$\Sigma$) the total mass would be 241 \MSun, and 55 \MSun for its core. 

However, the clusters likely contained an even larger star population in their infancy because a fraction of their stars spread in tidal tails at larger distances from the cluster centre than considered in this paper (e.g. \citealt{Roeser2019}). The tidal tails are structures in transition, becoming part of the galactic field and creating a link to galactic stellar population signatures. In the case of Hyades, \citet{Jerabkova2021} traced the tails up to a length of 800 pc, which is way beyond our definition of Solar neighbourhood ($<100$ pc). Therefore, a small fraction of these tails are considered for this study. Estimates of the original mass $M_{\rm Orig}$ of Hyades range from 800 to 1,600 \MSun \citep{Weidemann1992, Portegies2001, Kroupa2002}, which is a factor between 2 and 4 larger than what we estimated above. Recent determinations range from $>780$ \MSun considered by \citet{Oh2020} to 1721 \MSun obtained by \citet{Ernst2011}, with the determination of $\sim 1200$ \MSun obtained by \citet{Jerabkova2021} as an intermediate value. This implies that less than half of the stars that formed in Hyades are still recognised as associated with the cluster. This fraction may be even lower in the case of Ursa Major and Coma Berenices. We may also consider that the mass lost by open clusters is represented by the formula $d(M/M_\odot)/dt =-(M_{\rm Core}/M_\odot)^{1-\gamma}/t_0$ \citep{Lamers2010}, where $\gamma$\ depends on the initial concentration (with values in the range 0.6 - 0.8 being typical) and $t_0$ on the environment (which is similar for the three clusters considered here). If we assume that a value of $t_0=4$ Myr is typical for the solar neighbourhood \citep{Lamers2010}, we can calibrate the value of $\gamma$ assuming that the original mass of Hyades was $1200\pm 400$ \MSun, which represents the range considered above. Using this approach, we obtain $\gamma=0.765^{+0.109}_{-0.057}$. If we now assume the same value of $\gamma$\ for the other clusters, we estimated that the original mass of the Coma Berenices and UMa clusters was $587\pm 245$ and $415\pm 165$ \MSun, respectively. While these values are quite uncertain, we are here mainly interested in an order-of-magnitude estimate, and therefore we adopt them in our discussion. Since all of these clusters have masses at birth $>250$ \MSun, which is higher than the NYMGs discussed in \citet{Gratton2024}, they are more representative of the typical environment for star formation according to \citet{Winter2022b}.

\subsection{Metallicity}
\label{sec:metallicity}

Hyades are known to be slightly more metal rich than the Sun. Among a long list of abundance determinations, we consider here the accurate value of [Fe/H]=$0.13\pm 0.01$ obtained by \citet{Paulson2003}. This agrees with the average of the four stars considered by \citet{Valenti2005} ([Fe/H]=$0.13\pm 0.09$). A slightly higher value of [Fe/H]=$0.16\pm 0.02$ has been obtained by \citet{Liu2016b}. These studies also obtained high homogeneity among the different members of the cluster, although they cannot rule out some subtle differences \citep{Casamiquela2020}.

Coma Berenices (Melotte 111) and the Ursa Major group likely have a solar composition. For Coma Berenices various determinations are [Fe/H]=$-0.03\pm 0.04$ \citep{Cayrel1988}, [Fe/H]=$-0.05\pm 0.05$ \citep{Friel1992}, [Fe/H]=$0.07\pm 0.09$ \citep{Gebran2008}. In the case of the Ursa Major group \citet{Cayrel1988} gives [Fe/H]=$-0.02\pm 0.05$, \citet{Friel1992} [Fe/H]=$-0.06\pm 0.02$, \citet{Biazzo2012} [Fe/H]=$0.01\pm 0.01$, and if we consider the three stars in \citet{Valenti2005} we obtain [Fe/H]=$0.04\pm 0.04$.

\section{Detection of companions}
\label{sec:companion_detection}

\subsection{Visual binaries}

We searched for visual companions considering several sources, including:
\begin{itemize}
    \item Candidate companions with similar distance and proper motion within 600 arcsec listed as separate entries in the Gaia DR3 catalogue \citep{Gaia_DR3}.
    \item Visual binaries from \citet{Tokovinin2018}.
    \item Companions with small separation ($<2$ arcsec) in the WDS catalogue \citep{Mason2001}.
    \item Companions found within direct imaging surveys conducted with ESO high contrast imagers (SPHERE and NACO), GPIES \citep{Nielsen2019}, IDPS \citep{Galicher2016}, Gemini Deep Planet Survey \citep{Lafreniere2007}, \citet{Janson2011}, and LEECH \citep{Stone2018} surveys. 
    \item For Hyades and Coma Berenices clusters, we also considered speckle interferometry by \citet{Patience1998} and \citet{Guerrero2015}, respectively.
\end{itemize}

Companions discovered through speckle and HCI are so close to the stars that the probability that they are either field or unrelated cluster stars is low. On the other hand, several apparent wide companions selected from Gaia may be not bound to the primaries, but rather be separated members of the clusters projected close to the programme star. To distinguish between these cases, we estimated the probability that another member of the cluster is projected at a separation lower than that of the proposed companion to every star using a structural model for the cluster and considering the projected distance of the star from the cluster centre. Whenever this probability is higher than 1\% (50\%), we added the note 'possibly (likely) an unrelated cluster member'. For this purpose, we estimated the surface density in a region around each star if the cluster members follow a Plummer distribution. The appropriate parameters of the Plummer distribution for Hyades were from \citet{Roeser2011} and for Coma Berenices from \citet{Tarricq2022}. The surface stellar density of the Ursa Major cluster is much lower, so that all candidate companions projected within 600 arcsec of each star in this cluster can be considered as physically associated with the star. 

\subsection{Photometric binaries}

Companions whose orbital plane is close to the Sun can be discovered by eclipses or transits. We can detect only some companion this way. We searched for entries corresponding to the programme stars in the eclipsing binary (EB) catalogue by \citet{Avvakumova2013}, in the Kepler EB catalogue \citep{Slawson2011, IJspeert2024}, and in the NASA Transiting Exoplanet Survey Satellite (TESS) catalogues by \citet{Prsa2022} and \citet{IJspeert2021} . All 353 programme stars but 8 have been observed by the TESS satellite in short-cadence mode and then have appropriate light curves; an additional four were not included in the input catalogue but have been observed in the long-cadence mode. The stars missing TESS data are Sirius because of its brightness, and HIP 23312, TYC0078-257-1, and 2MASS J12485421+1407213 because they fall in CCD gaps.

This search returned five EBs among the programme stars: HIP 20019 and HIP 24019 in Hyades \citep{Avvakumova2013}; HIP 61071 in Coma Berenices \citep{Prsa2022}; HIP 28360 (beta Aur) and HIP 76267 (alpha CrB) in Ursa Major tail \citep{Avvakumova2013, IJspeert2021, Prsa2022}.

\subsection{Spectroscopic binaries}

We considered the information we could find in the literature. Time series of radial velocities (RV) of moderate precision are available from Gaia for a large fraction of the stars. Many stars in Hyades and in the other clusters were searched for spectroscopic binaries (see references for individual stars in the Appendix~\ref{appendix:data_tables}). In addition, we looked for high-precision RVs in archives that are available on-line: Keck \citep{Tal-Or2019}; HARPS \citep{Trifonov2020}; SOPHIE and ELODIE \citep{Perruchot2008}\footnote{\url{http://atlas.obs-hp.fr/sophie/}, \url{http://atlas.obs-hp.fr/elodie}}; \citet{Grandjean2021}; \citet{Paulson2006}; \citet{Quinn2016}. In total, we found high-precision RV series for a total of 167 stars (mainly in Hyades, thanks to devoted surveys such as that from \citealt{Cochran2002}), and low-precision RV series for 152 additional stars, which are in total 90.3\% of the stars. Most stars missing these data have close visual companions whose contamination prevents accurate RV measurements from Gaia.

\subsection{Astrometric binaries}

We inspected the Gaia $gaiadr3.nss\_two\_body\_orbit$ catalogue \citep{2023A&A...674A..34G} and found entries for 19 objects: HIP59974 and HIP80686 in Ursa Major; 2MASS J12075772+2535112, 2MASS J11562380+3038284, HD106293, HIP71657, HIP69275, and HIP72389 in Coma Berenices; HIP18658, HD281459, HIP20419, HIP20482, HIP20614, HIP21029, HD28545, HIP21123, HIP21273, HIP22203, and HIP22565 in Hyades. An orbit solution is available for these stars. We found 9 entries in the $gaiadr3.nss\_acceleration\_astro$ catalogue \citep{2023A&A...674A..34G}: HIP8486 in Ursa Major; 2MASS J14022156+5225039 in Coma Berenices; HD28034, HIP20719, HD285766, HIP21112, HD29896, HIP22265, and HIP22350 in Hyades. We searched for the programme stars in the catalogue of astrometric orbit determinations with Markov chain Monte Carlo and genetic algorithms by \citet{Holl2023} and found only one entry (HIP22203 in Hyades).

We considered Proper Motion Acceleration (PMa) from \citet{Kervella2022}. The PMa is the difference between the proper motion in Gaia DR3 (baseline of 34 months) and that determined using the position at Hipparcos (1991.25) and Gaia EDR3 (2016.0) epochs. This quantity is available for 247 stars. The PMa is sensitive to binaries with a projected separation between 1 and 100 au. We considered any value of PMa with a signal-to-noise ratio $SNR>3$ as an indication of the presence of companions. 

We also considered the re-normalised unit weight error (RUWE) as an indication of binarity. This parameter is an indication of the goodness of the 5-parameter solution found by Gaia \citep{Lindegren2018}. \citet{Belokurov2020} showed that a value $>1.4$ of this parameter is an indication of binarity, at least for stars that are not too bright ($G>4$) and saturated in the Gaia scans. This method is sensitive to systems that have periods from a few months to a decade \citep{Penoyre2021}. The RUWE parameter is available for all but 21 stars, which means for 94\% of the cases. We considered these data in our analysis.

\section{Companion parameters}
\label{sec:companion_parameters}

\begin{figure}[htb]
    \centering
    \includegraphics[width=8.5cm]{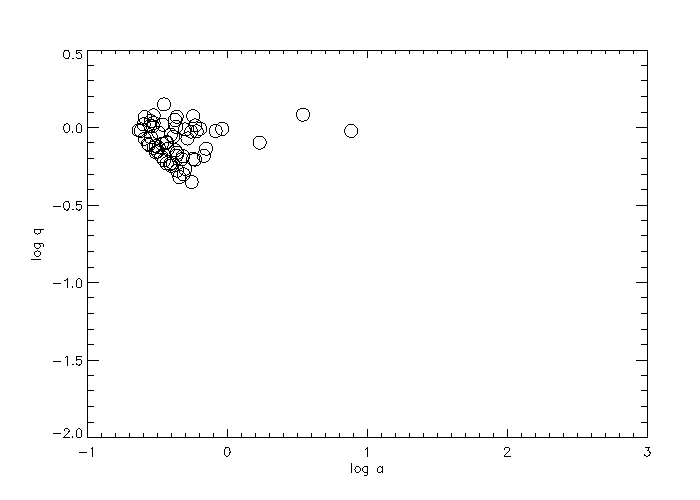}
    \includegraphics[width=8.5cm]{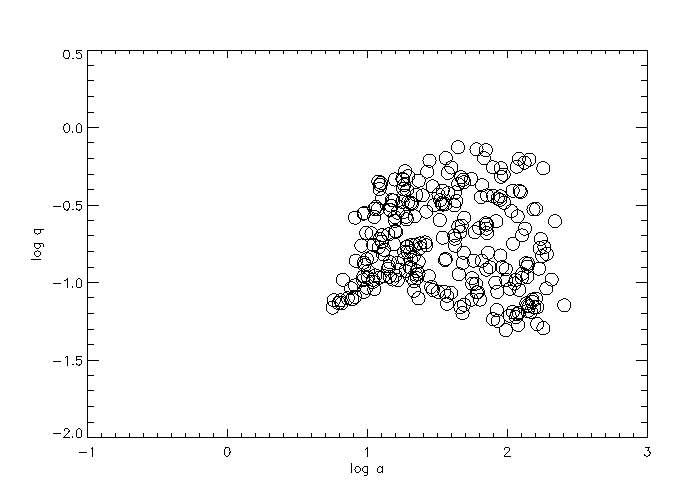}
    \caption{Distribution of semi-major axis $a$ and mass ratios $q$ for companions producing values of $RUWE$, scatter in RVs, and $SNR(PMa)$ compatible with the observed values for two stars within our sample. The upper panel gives the results for TYC 1986-852-1 (where the indication of the presence of a companion is given by the high value of $RUWE$); the lower panel for HIP 59833 (where the indication of the presence of a companion is given by the high value of $SNR(PMa)$). The adopted final values are $\log{a/au}=-0.169\pm 0.189$ and $\log{q}=-0.125\pm 0.103$, and $\log{a/au}=1.32\pm 0.34$ and $\log{q}=-0.66\pm 0.23$ for TYC 1986-852-1 and HIP 59833, respectively.}
    \label{fig:two_cases}
\end{figure}

\begin{figure}[htb]
    \centering
    \includegraphics[width=8.5cm]{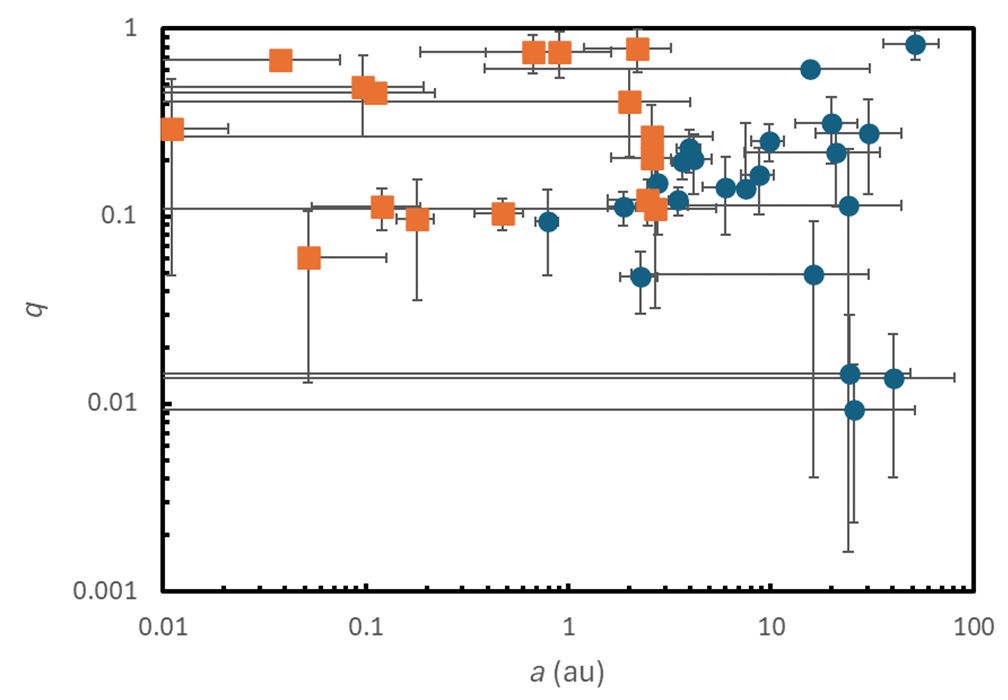}
    \caption{Relation between semi-major axis $a$ and mass ratio $q$ for the companions of the intermediate-age cluster stars discovered through Gaia PMa, RUWE, or variations in RVs that are not visual binaries or do not have an orbit determination. Orange filled squares are companions selected on the basis of the value of $RUWE>1.4$; blue filled circles are those selected from SNR(PMa)$>3$. }
    \label{fig:dyn}
\end{figure}

\begin{table*}
\caption{Masses and separations derived using dynamical information and photometric upper limits.}
\scriptsize
\begin{tabular}{lccccccc}
\hline
HIP	&	HD	&	2MASS	&	Others	&	$M_A$	&	$M_B$	&a	&Selection\\
	&		&		&		&	\MSun	& \MSun	&	au	&\\
\hline
\multicolumn{7}{c}{Ursa Major Core}\\											\hline
59496	&238087	&J12120522+5855351&BD+59 1428	&	0.72&$0.120\pm 0.046$	&$8.78\pm 1.64$	& PMA \\
\hline
\multicolumn{7}{c}{UMA other members}\\
\hline
21276	&	28495	&	J04335424+6437593	&	MS Cam 	&	0.95&$0.192\pm 0.067$	&$4.17\pm 0.97$	& PMA \\
43352	&75605	&J08495150-3246497&HR3512	&	2.91&$0.356\pm 0.062$	&$3.49\pm 0.35$	& PMA \\
77233	&141003	&J15461125+1525185&bet Ser	&	2.96&$0.141\pm 0.051$	&$2.28\pm 0.48$	& PMA \\
88694	&165185	&J18062370-3601113&		&	1.04&$0.146\pm 0.182$	&$7.50\pm 0.18$	& PMA \\
93408	&177196	&J19012637+4656055&16 Lyr	&	1.78&$0.493\pm 0.258$	&$30.37\pm 13.81$	& PMA \\
103738	&199951	&J21011746-3215281&gam Mic	&	2.91&$0.733\pm 0.164$	&$9.80\pm 1.81$	& PMA \\
113136	&216627	&J22543901-1549144&del Aqr	&	3.22&$0.486\pm 0.230$	&$2.77\pm 0.068$	& PMA \\
\hline
\multicolumn{7}{c}{Coma Core}\\
\hline
	&	&J12075772+2535112&TYC 1986-852-1&	0.88&$0.66\pm 0.154$	&$0.67\pm 0.277$	& RUWE \\
	&	&J12113516+2922444&TYC 1988-920-1&	0.88&$0.40\pm 0.036$	&$0.111\pm 0.111$	& RUWE \\
59833	&106691	&J12160837+2545373&		&	1.41&$0.31\pm 0.15$	&$21.087\pm 13.69$	& PMA \\
60206	&107399	&J12204557+2545572&		&	1.09&$0.67\pm 0.027$	&$15.718\pm 15.33$	& PMA \\
60304	&	107611	&	J12215616+2718342	&			& 1.22 &$0.060\pm 0.055$	&$16.2918\pm 14.24$	& PMA \\
60406	&107793	&J12230840+2551049&		&	1.09&$0.317\pm 0.267$	&$0.011\pm 0.018$	& RUWE \\
60458	&107877	&J12234101+2658478&		&	1.29&$0.012\pm 0.009$	&$25.7\pm 36.99$	& PMA \\
60611	&108154	&J12252249+2313447&		&	1.22&$0.14\pm 0.138$	&$24.118\pm 20.21$	& PMA \\
	&	&J12573686+2858448&BD+29 2346	&	0.97&$0.26\pm 0.119$	&$2.573\pm 3.041$	& RUWE \\
\hline
\multicolumn{7}{c}{Coma Tails}\\
\hline
	&94041	&J10514921+3225555&		&	1.06&$0.13\pm 0.036$	&$2.472\pm 0.896$	& RUWE \\
	&	&J13211119+3748487&BD+38 2436	&	1.07&$0.12\pm 0.030$	&$0.12\pm 0.066$	& RUWE \\
70051	&25658	&J14200866+3025449&HR 5374	&	1.87&$0.21\pm 0.043$	&$1.869\pm 0.302$	& PMA \\
71657	&129046	&J14392466+3630378&		&	1.28&$0.12\pm 0.058$	&$0.79\pm 0.102$	& PMA \\
\hline
\multicolumn{7}{c}{Latishev 2}\\
\hline
67005	&119765	&J13435477+5203520&		&	2.29&$0.138\pm 0.108$	&$0.052\pm 0.074$	& RUWE \\
	&	&J15503671+5250220&TYC 3870-1328-1&	1.07&$0.22\pm 0.064$	&$2.58\pm 0.941$	& RUWE \\
77652	&142282	&J15510936+5254259&		&	1.97&$0.385\pm 0.075$	&$3.664\pm 0.404$	& PMA \\
\hline
\multicolumn{7}{c}{Mecayotl 1}\\
\hline
	&	&J13451646+5214502&TYC 3470-485-1&	0.84&$0.633\pm 0.175$	&$0.906\pm 0.719$	& RUWE \\
	&	&J14022156+5225039&TYC 3471-333-1&	0.96&$0.106\pm 0.075$	&$2.683\pm 2.679$	& RUWE \\
	&	&J14570087+5833289&TYC 3867-1319-1&	0.79&$0.622\pm 0.161$	&$2.20\pm 1.006$	& RUWE \\
\hline
\multicolumn{7}{c}{Hyades}\\
\hline
19796	&26784	&J04143433+1042050&vB 19	&	1.23&$0.017\pm 0.012$	&$40.157\pm 142.65$	& PMA \\
20255	&	27429	&	04202508+1844334	&	vB 32 & 1.54 &$1.85\pm 0.23$ & $51.75\pm 15.59$ & PMA \\
20400	&27628	&J04220349+1404379&60 Tau, vB38	&	1.64&$1.122\pm 0.108$	&$0.038\pm 0.038$	& RUWE \\
20484	&27749	&J04232504+1646382&63 Tau, vB45	&	1.72&$0.847\pm 0.390$	&$0.097\pm 0.097$	& RUWE \\
21112	&28635	&J04312933+1354125&		&	1.11&$0.256\pm 0.066$	&$3.977\pm 0.547$	& PMA \\
21459	&29169	&J04362913+2320270&V1116 Tau, vB 100&	1.51&$0.217\pm 0.097$	&$5.983\pm 1.345$	& PMA \\
21474	&29225	&J04364070+1552094&vB 101	&	1.38&$0.431\pm 0.168$	&$19.942\pm 6.689$	& PMA \\
21637	&29419	&J04385127+2309000&vB 105	&	1.10&$0.016\pm 0.017$	&$24.470\pm 42.076$	& PMA \\
21683	&29488	&J04391649+1555048&sig02 Tau, vB 108&	2.41&$0.252\pm 0.049$	&$0.473\pm 0.129$	& RUWE \\
22157	&30210	&J04460173+1142200&vB 112	&	2.19&$0.212\pm 0.134$	&$0.180\pm 0.038$	& RUWE \\
	&	&J05090304+1257510&BD+12 736	&	0.92&$0.375\pm 0.185$	&$2.011\pm 2.060$	& RUWE \\
\hline
\end{tabular}
\label{tab:dyn}
\end{table*}

With the approach described in Section \ref{sec:companion_detection} we found 55 companions in 46 of the 64 systems considered in the Ursa Major association and tails, 74 companions over 57 of the 119 systems considered in the Coma Berenices cluster and tails, and 146 companions over 102 of the 169 systems considered in Hyades cluster and tails. While other properties of the companions (e.g. the eccentricity of their orbits) are of high interest to determine their origin, in this paper we only considered the semi-major axis of the orbit and the mass of the companion because information about eccentricity could be derived only for a small fraction of the objects. Overall, we derived estimates of the mass and semi-major axis for almost all companions. We did not have an estimate of the semi-major axis for the close companion to 2MASS J15042576+5952507 (BD+60 1587), whose evidence is clear from the colour excess, but it is not resolved, and it has not enough photometric and spectroscopic data. We do not have adequate information to derive an estimate of the mass of the close companion to HIP18735. Even though our estimates of the semi-major axes and masses have quite large uncertainties in a significant fraction of the cases, they are still useful for the statistical discussion done in this paper. 

Whenever possible, the estimates of the semi-major axis and mass were taken from sources in the literature listed in Table \ref{tab:mass_ursa_major}. When this was not possible, we followed the methods considered in \citet{Gratton2023, Gratton2024}. They are briefly summarised in the following.

We derived the masses of visual binaries from the photometry for the individual stars and the isochrones by \citet{Baraffe2015}. We assumed that the semi-major axis is equal to the projected separation divided by the parallax. On average, this corresponds to the thermal eccentricity distribution considered by \citet{Ambartsumian1937} of $f(e)=2 e$ (see \citealt{Brandeker2006}). This last paper indicates that this assumption underestimates the semi-major axis by about 25\% in the case of circular orbits. This can be compared with the eccentricity distribution for wide binaries by \citet{Hwang2022}. These authors found that the eccentricity distribution ranges from uniform (that yields a mean value of $e=0.5$) for $a<100$ au, to thermal (that yields a mean value of $e=0.67$) and even suprathermal for $a>300$ au. Uncertainties in the masses derived using these recipes are small (well below 10\%), while those for the semi-major axes are about 40\% (see Figure A.1 in \citealt{Brandeker2006}).

For double-lined spectroscopic binaries, whenever possible the mass ratios were derived from the ratio of the amplitudes of the spectroscopic orbits, and the sum of the masses was made compatible with the apparent $G$ magnitude of the system, using the mass-luminosity relation for the Gaia $G$ band by \citet{Pecaut2013}. For several single-lined spectroscopic orbits, we obtained the masses from the amplitude of the primary motion, assuming circular orbits, a median value of the distribution of inclinations (60 degrees), and again a sum of the masses compatible with the apparent luminosity of the system. We obtained the semi-major axis for these binaries from the periods and the masses using the Kepler third law. In this case, while the uncertainties in the semi-major axis are small, those on the masses may be substantial in those cases where the orbit inclination is small.

The indication of binarity comes from PMa, RUWE, or RVs for 38 objects. The secondary of these stars was not observe directly and no period or semi-major axis was determined. These binaries have a different range of semi-major axes, depending on the technique used to detect them (see Figure 3 of \citealt{Gratton2024}). These plots indicate that binaries discovered through RVs have semi-major axes that are typically lower than about 1 au, those discovered through $RUWE$ have semi-major axes in the range 1-10 au, while those detected through PMa have a typical semi-major axis of around 10 au. In fact, when considering astrometry, the semi-major axis for which maximum sensitivity is achieved depends on the length of the baseline considered: this is 34 months for Gaia DR3 observations used for the case of $RUWE$, and 24.75 yr for the separation between the Hipparcos and Gaia DR3 epochs used for SNR(PMa). Of course, we should also consider the typical total mass of these binaries (about 1.5 \MSun). The different sensitivities of the various techniques indicate that we may better constrain masses and separations combining the different methods rather than considering only a single technique.

In Table \ref{tab:dyn} we list objects that have indications for the presence of companions from PMa, RUWE, or RVs. For these objects, we obtained solutions that are compatible with the observed values of RUWE and PMa, the scatter in RV, and the lack of detection in HCI (if available, else in Gaia), as done in \citet{Gratton2023, Gratton2024}. We did this exploring the semi-major axis - mass ratio plane using a Monte Carlo code. We adopted eccentric orbits, with uniform priors between 0 and 1 in eccentricity (which is in agreement with \citealt{Hwang2022} for this range of separations), 0 and 180 degrees in the angle of the ascending node $\Omega$, and 0 and 360 degrees in the periastron angle $\omega$, and left the inclination and phase to assume a random value. In addition, the period was used to fix the solution whenever it was available. Figure \ref{fig:two_cases} shows the distribution of the acceptable solutions obtained using this procedure in the $a-q$ plane for two typical cases, one where the indication of binarity is driven by the high value of $RUWE$ and the other by the high value of $SNR(PMa)$. The values we finally adopted are the mean of those obtained for solutions compatible with observations within the errors, and the uncertainty is the standard deviation of this population. Data for the derivation of these masses are given in Table \ref{tab:binary_ursa_major} in Appendix~\ref{appendix:data_tables}; the values of $a$ and masses of the stars are given in Table \ref{tab:dyn}. The distribution of the companions in the $a-q$ plane is shown in Figure \ref{fig:dyn}; they have $0.01<a<100$ au, with companions whose presence is indicated by a high value of $RUWE$ having on average higher values of $q$ and lower values of $a$ than those whose presence is indicated by a high value of PMa. The median errors in the masses and semi-major axes derived by this method are 39\% and 43\%, respectively.

\section{Substellar companions}
\label{sec:substellar_companions}

In our study we found a dozen stars with substellar companions, some of them already known. The derivation of companion parameters follows the same methods described in Section \ref{sec:companion_parameters}. In the following, we give some details about them. We remark that other members of these three groups, less massive than the stars here considered, are known to host planets (e.g. K2-25, \citealt{Stefansson2020} and K2-136, \citealt{Mann2018, Mayo2023} in Hyades with masses of 0.26 \MSun and 0.74 \MSun and one and three transiting planets, respectively).

\subsection{Ursa Major}

The close companion of HIP14844 (mass of 1.81 \MSun) in the Ursa Major moving group may be a BD with a nominal mass (assuming an inclination of 60 degrees) of 0.07 \MSun and $a=0.12$ au \citep{Tokovinin2018}.

HIP38228 (mass of 1.00 \MSun) in the tail of the Ursa Major association was found to host three sub-Neptune planets from TESS and high-precision radial velocities \citep{Damasso2023, Capistrant2024}. However, there is no obvious sign for the presence of more massive companions.

HIP88694 (mass of 1.04 \MSun) in the tail of the Ursa Major association has a small but significant PMa value (SNR=3.31: \citealt{Kervella2022}), and not significant value of RUWE. The low-mass (0.17 \MSun) wide companion at 12.28 arcsec (projected distance of 210 au) is unlikely to be the cause of the astrometric perturbation. The star has been observed in HCI \citep{Desidera2021} with no detection of a massive companion. It might then host a JL planet. The star is very likely a member of the Ursa Major group.

HIP94083 (mass of 1.04 \MSun) in the tail of the Ursa Major association has a close companion that probably is a BD with a mass of 0.025 \MSun and $a=0.21$ au as indicated by high-precision RVs \citep{Galland2006}.

\subsection{Coma Berenices}

HIP60458 (mass of 1.29 \MSun) in the core of the Coma Berenices cluster may possibly have two warm Jupiter planets ($P=43.8$ d, $M \sin{ i}=2.49$ M$_{\rm Jup}$; $P=450$ d, $M \sin{ i}=2.4$ M$_{\rm Jup}$: \citealt{Quinn2016}). This star has significant RUWE and PMa, which when interpreted as a single companion are compatible with a planet with a mass of $12\pm 9$ M$_{\rm Jup}$ and $a=25.7^{+37.0}_{-25.7}$ au. In either case, the star is likely to host a JL planet.

2MASS J15514181+5218226 (TOI-2048; mass of 0.91 \MSun) in Mecayotl 1 hosts a transiting planet discovered in TESS data with $P=13.79$ d and $R=2.6$ R$_{\rm Earth}$ \citep{Newton2022}.

\subsection{Hyades}

HIP19796 (mass of 1.23 \MSun) at 5.3 pc from Hyades cluster centre has a small but significant PMa value (SNR=3.09: \citealt{Kervella2022}) but not a significant value of RUWE. The companion has not been detected so far in HCI, though a search in the archive showed that it has been observed with SPHERE. Since a stellar secondary of a massive BD companion would be detectable in such images, the star might host a massive planet or a low-mass BD. The solution we obtain combining RUWE, RV, and PMa indicates a mass of $17\pm 12$ M$_{\rm Jup}$ and a semi-major axis of the order of 40 au (see Section~\ref{sec:companion_parameters}). We hereafter assume that this companion is a JL planet.

HIP20889 ($\epsilon$ Tau) at 2.9 pc from Hyades cluster centre is one of Hyades red giants, with a mass of 2.7 \MSun \citep{daSilva2006}. The WDS catalogue \citep{Mason2001} lists a close visual stellar companion obtained using speckle interferometry \citep{Mason2009}, but this is not compatible with the existing RV and the value of the PMa. The high-precision RVs by \citet{Teng2023} rather indicate the presence of a giant planet with $M \sin{ i}=7.19$ M$_{\rm Jup}$ with $a=1.9$ au. We hereafter assume that this companion is a JL planet.

HIP21152 (mass of 1.41 \MSun) at 9.3 pc from Hyades cluster centre (i.e., it is just outside the tidal radius) is known to host a low-mass BD companion with a mass of $M=24$ M$_{\rm Jup}$ and a projected separation of 0.4 arcsec, corresponding to 17 au \citep{Bonavita2022, Kuzuhara2022, Franson2023c} detected in HCI, after the object was selected for observation in view of the high value of the PMa (SNR=10.66: \citealt{Kervella2022}). Although the mass of this companion is higher than the planetary limit, the mass ratio $q=0.017$ is below the upper limit considered for JL planets according to the definition considered in this paper.

HIP21637 (mass of 1.10 \MSun) at 4.7 pc from Hyades cluster centre has a small but significant PMa value (SNR=4.04: \citealt{Kervella2022}) but not a significant value of RUWE. The companion has not been detected so far in HCI, though search in the archives shows that it has been observed with SPHERE. Since a stellar secondary of a massive BD companion would be detectable in such images, we suggest that the star might rather host a massive planet or a low-mass BD. The solution we obtain by combining the data indicates a mass of $16\pm 16$ M$_{\rm Jup}$ and $a\sim 24$ au (see Section~\ref{sec:companion_parameters}). We then hereafter assume that this companion is a JL planet.

HIP22203 (mass of 1.02 \MSun) at 2.9 pc from Hyades cluster centre is a spectroscopic binary whose orbital solution ($gaiadr3.nss\_two\_body\_orbit$, \citealt{Bender2008, Diaz2012} is compatible with a low-mass star or a BD, with a nominal mass (assuming an inclination of 60 degrees) of 0.055 \MSun and $a=2.01$ au.

HIP24019 (a mass of 1.83 \MSun) at 23.7 pc from Hyades cluster centre (which is well further than the tidal radius) is claimed to be an EB in the Simbad database, but there is no sign of an eclipse in the TESS data. However, the star is an SB1 with a period of 32.522 d and an amplitude of 2.5 \kms \citep{Tokovinin2018, Debernardi2000}. This is compatible with a low-mass star or a BD, with a nominal mass (assuming an inclination of 60 degrees) of 0.035 \MSun and $a=0.246$ au. 

\subsection{Controversial cases}

HIP61295 in Coma Berenices: \citet{Abt1999} suggests that this star is an SB with a period of 175.2 d; however, the Gaia RV looks constant.

HIP69275 (2MASS J14105307+6231202, mass of 1.40 \MSun) in Latishev 2, at 102 pc from the Sun. The star has significant RUWE and PMa and a variation of RV in Gaia. It is signalled as an SB in Gaia with a period of 195.8643 d. RUWE and variations in RV are well consistent with a companion of around 0.48 \MSun at 0.81 au but PMa is not expected to be significant for such a compact system. However, the star has a companion at a separation of 832 mas and PA=352.647 degree with a contrast of about 6.59 mag from WIYN speckle imaging using NESSI (wavelength of 832 nm\footnote{See \url{https://exofop.ipac.caltech.edu/tess/edit_obsnotes.php?id=166053959}.}), which corresponds to star with a mass of 0.29 \MSun at about 85 au. This object is likely responsible for the observed PMa. In addition, the star is also a TESS object of interest (TOI-5383.01) with a period of 2.81 d; if this signal is due to a planet transiting the primary disk, the object that causes the photometric dip of 90 ppm has a radius of 1.4 R$_{\rm Earth}$. An alternative explanation is that the photometric dips observed by TESS are on one of the two companions. However, in both cases they would be planets with radii of about 6-7 R$_{\rm Earth}$, which is in the middle of the so-called Neptune desert \citep{Szabo2011}. It is then more probable that the planetary object is around the primary and it is not a JL planet.

\section{Search completeness}
\label{sec:completeness}

\begin{figure*}[htb]
\centering
\includegraphics[width=\linewidth]{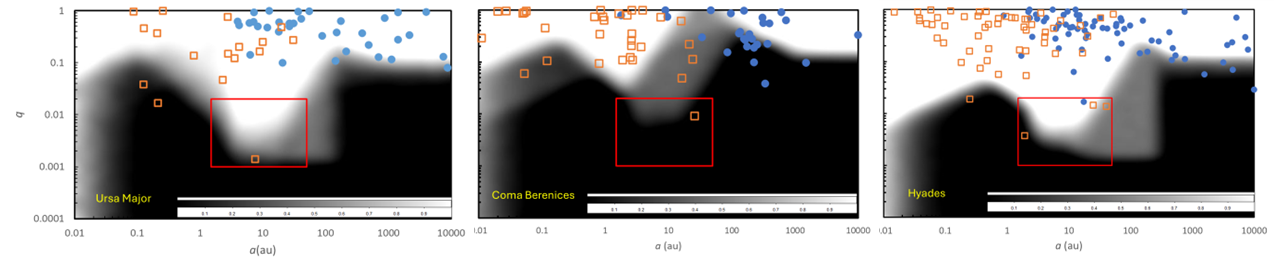}
\caption{Relation between semi-major axis $a$ (in au) and mass ratio $q$ between components for all companions found so far around stars that are member of the nearby intermediate-age open clusters considered in this paper. Blue filled circles are resolved companions (detected in HCI or directly seen in Gaia); orange open squares are those detected using dynamical methods or eclipses or transits. Detection completeness in background; a linear grey-scale is used with white meaning 100\% and black zero (see the insert on bottom right of each panel). The red rectangle marks the area covered by JL planets, here $1.5<a<50$ au and $0.001<q<0.02$. Left panel is for Ursa Major; centre panel is for Coma Berenices; right panel is for Hyades.  Candidate companions with a probability higher than 50\% of being unrelated cluster members were not considered in these plots.
}
\label{fig:detection_vs_completeness}
\end{figure*}

\begin{table}
\centering
\caption{Summary of available data.}
\begin{tabular}{lcccc}
\hline
Parameter    &  Ursa Major  &  Coma  &  Hyades & Total \\
\hline
Tot. systems      & 64 & 119 & 170 & 353 \\
In TESS/Kepler2   & 60 & 117 & 168 & 345 \\
High-precision RV &  6 &  29 & 132 & 167 \\
Low-precision RV  & 43 &  77 &  32 & 152 \\
HCI               & 28 &   0 &  61 &  89 \\
RUWE              & 52 & 118 & 162 & 332 \\
PMa               & 54 &  51 & 142 & 247 \\
\hline
\end{tabular}
\label{tab:data_summary}
\end{table}

\subsection{Data used}

Table \ref{tab:data_summary} summarises the data we used in the search for companions. Though wide visual binaries are detected around all stars as separate entries by Gaia, only about a quarter of the stars have HCI, none in Coma. RUWE is available for nearly all stars (data is missing only for a few close visual binaries). Photometric series from TESS and or Kepler2 are available for all but eight programme stars. The search for eclipsing binaries and transiting hot giant planets should then be quite complete. RV series are also available for more than 90\% of the stars, but only a minor fraction of the stars of the Ursa Major and Coma Berenices clusters have high-precision data. This implies that while the search for spectroscopic binaries is nearly complete, only about half of the stars were searched for hot and warm planets. Finally, PMa is available for slightly more than 70\% of the stars. This means that we should have detected most binaries, while there is some (though limited) sensitivity to JL planets.

\subsection{Completeness maps}

Once the availability of data is known, we may use it to estimate the completeness of our search for companions as a function of the semi-major axis $a$ and the mass ratio $q$ considering the detection limits appropriate for each method used to search for companions. This is illustrated in Figure \ref{fig:detection_vs_completeness} for each individual cluster. These completeness maps were obtained using the same approach described in \citet{Gratton2023} with updates to make it more appropriate for solar-type stars as described in \citet{Gratton2024}. The code used here is identical to that used in that paper. The reader may find a full description of this procedure in those papers; this procedure is also similar to that proposed by \citet{Wood2021}. Briefly, we simulated 10,000 companions (stellar and planetary) for each of the stars in our sample (with appropriate values of the magnitude, mass, parallax and age) with random values of the semi-major axis $a$, mass ratio $q$, inclination $i$ and phase, assuming circular orbits (this choice is commented on in \citealt{Gratton2023}). We considered uniform distributions in the logarithm for $a$ and $q$ (with $0.01<a<10000$ and $0.0001<q<1$), uniform distributions of phase between 0 and 1, and isotropic distributions of inclinations. For each target and methodology (visual, eclipsing, spectroscopic, and astrometric binaries), we considered whether the appropriate data sets were available for the star considered. We then determined the signal expected for each simulated companion according to the various methods considered (transit/eclipses, RVs, Gaia imaging/HCI, RUWE, PMa) and if the datum is available for the considered star, we compared this signal with the detection limits appropriate for each method. We considered that each simulated companion that produces a signal above the detection limits is detected. In order to reduce local fluctuations due to low number statistics, we finally smoothed the detection distribution in the ($\log{a}$ and $\log{q}$) plane with a Gaussian kernel of 0.1 dex; this ensures that of the order of a few hundred simulated planets are considered for each position in the ($\log{a}$ and $\log{q}$) region explored by this simulation. 

\subsection{Detection limits}

Although the detection limits are possibly questionable, they were used consistently in the simulation and in the real data. 

The limiting contrasts of the HCI are as in \citet{Gratton2024}. The limiting contrast of Gaia DR2 is discussed in \citet{Brandeker2019}; however, we found that many faint companions at separation of a few arcsec present in Gaia DR3 have contrasts that yield a very low detection probability according to the table of that paper, and should not be detected using the 50\% detection limits (see Appendix~\ref{appendix:dr3_detlim}). Rather, we find that the 1\% detection limits there considered better reproduce observations and attribute this difference to the progresses achieved in DR3 with respect to DR2. 

For eclipses or transits the detection procedure is more complex and there is no simple threshold value available in the literature. Considering the stellar companions, we simply assumed that all EBs with periods $<28$ days are recovered by TESS. TESS is in principle also able to recover some EB with longer periods depending on the number of sectors where the target was observed. However, the catalogues of EB we used only refer to the early observations with TESS and very few stars were observed on many sectors. For transiting planets, the situation is much more complex and TESS is less efficient in discovering young transiting planets (e.g. \citealt{Fernandes2022, Fernandes2023}).

In the case of RV observations, the precision achievable for solar-type stars is in principle much higher than for early-type stars. This is due to the slower rotation of the stars and the much higher number of spectral lines available in their spectra. On the other hand, we should consider the concern related to the jitter due to stellar activity. The estimate of the RV jitter followed the precepts described in Appendix A of \citet{Gratton2024}. Adopting these thresholds, the completeness level for separation in the range 0.1-1 au is higher than obtained for the B-stars in \citet{Gratton2023}. We note that for what concerns eclipses, the detection efficiency that we consider preparing this figure includes the probability that the transit occurs.

The detection limits for the detection of astrometric binaries are $RUWE>1.4$ and $SNR(PMa)>3$, as in \citet{Gratton2023, Gratton2024}.

\subsection{Results using various methods}

Thanks to Gaia, the search for stellar visual companions should be complete down to a projected separation of a few hundred au. However, owing to the old age of the targets, most substellar companions are instead not detectable by Gaia. The masses corresponding to a magnitude of $G=19$ (which is the magnitude limit for stars with reasonably accurate astrometry) for the age and average distance of each association are listed in Table \ref{tab:associations}. They are in the BD regime, though close to the minimum stellar mass. HCI imaging, available for 89 stars, should reveal all stellar companions with semi-major axis in the range 5-150 au.

TESS data on the photometric variations are available for 345 stars. The search for eclipsing binaries with $a<0.2$ au (roughly a period of 27 days) should then be almost complete. However, we expect that only about 1/10 of the close binary systems should be eclipsing or transiting.

RV variations with clearly significant amplitude would be detected for $a<1$ au and for many others with larger radii. High-precision RV sequences are available for 167 stars, while RV series from Gaia are available for 152 additional stars in the sample.

Stellar companions with a separation in the 0.2-10 au range would produce a value of $RUWE>1.4$ and should be detected. RUWE is available for all but 21 stars in the sample. The PMa (available for 247 stars of the programme stars) should have $SNR>3$ for all systems that have a stellar companion in the 3-100 au separation range. There are 106 stars for which neither PMa nor HCI is available. For these stars (30\% of the programme stars) there may be incompleteness in the 10-100 au region. 

\subsection{Summary about completeness}

\begin{figure}[htb]
\centering
\includegraphics[width=\linewidth]{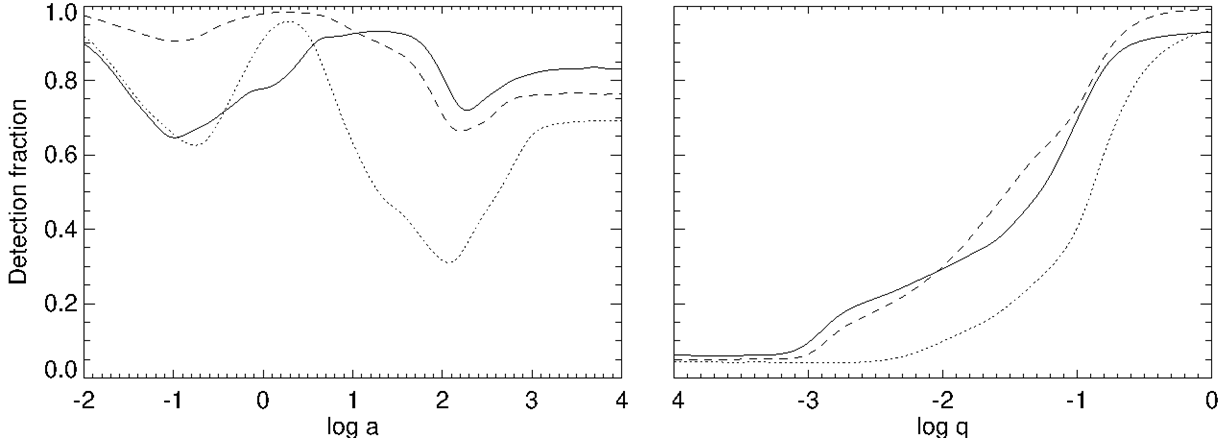}
\caption{Average completeness of companion search as a function of semi-major axis $a$ (left panel) and mass ratio $q$ (right panel). The upper panel is for stellar companions alone ($0.05<q<1$); the lower panel is computed over the whole range of separations considered in this paper ($-2<\log{a/au}<4$). In both panels, the solid line is for Ursa Major, the dotted line is for Coma Berenices, and the dashed line is for Hyades. 
}
\label{fig:completeness_a_q}
\end{figure}

\begin{figure*}[htb]
\centering
\includegraphics[width=\linewidth]{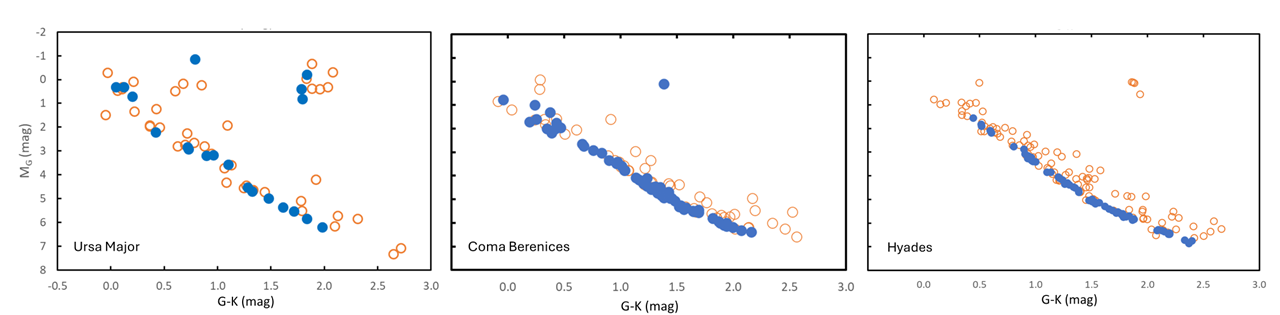}
\caption{Colour-magnitude diagrams of $M_G$ versus $G-K$ for the three open clusters. Blue filled circles are stars considered to be single or with a secondary at large separation ($>1000$ au). Open orange circles are stars with companions closer than this limit. The left panel is for Ursa Major; the centre panel is for Coma Berenices; the right panel is for Hyades. 
}
\label{fig:cmd}
\end{figure*}

The detection maps given in Figure \ref{fig:detection_vs_completeness} are obtained combining those obtained using individual methods: whenever a simulated companion produces a signal that is higher than the detection threshold for at least one of the search methods considered, and the relevant data are available for the star, the companion is assumed to be detectable. Figure 6 of \citealt{Gratton2024} provides an illustration of how this combination is made. Figure \ref{fig:completeness_a_q} gives the average detection fraction as a function of $a$ and $q$ for the different clusters. In the case of $a$, we considered the average detection fraction only for the range appropriate for stellar companions ($q>0.05$). 

Inspection of Figures \ref{fig:detection_vs_completeness} and \ref{fig:completeness_a_q} allowed us to conclude that while some massive companions may still be missing, we think our search is almost complete for stellar companions in Hyades and Ursa Major group, while some half of the stellar companions with semi-major axis between 10 and a few hundred au may be missing for the Coma Berenices cluster. The undetected companion frequency depends not only on efficiency of detection, but also on the actual mass ratio distribution of companions.

Although completeness in the binary search is on average high, we find indications that a few binaries may still be missing. Figure \ref{fig:cmd} shows the colour-magnitude diagram of $M_G$ versus $G-K$ for stars in the intermediate age clusters considered in this paper. We used different symbols for bona fide single stars and multiple stars. The single-star sequences are generally very narrow. However, a few binaries with low-mass companions may still be present among the sample of bona fide single stars. Furthermore, the search is less complete around stars that are more distant from the Sun. In fact, there is a trend for the fraction of single stars to decrease with the distance of the individual associations. This trend is due to the lack of PMa data for stars not included in the Hipparcos catalogue; in fact, the fraction of stars that have PMa data changes systematically with distance of the association, from more than 80\% for the closest cluster (Ursa Major) to about 40\% for the furthest one (Coma). This highlights the importance of the distance limit in our analysis.

\section{Discussion of results for stellar binaries}
\label{sec:discussion_stellar}

In this Section we discuss the results we obtained for stellar binaries in these intermediate age clusters. We consider separately the distributions of stellar companions with semi-major axis and mass, and then their frequency. For each of these features, we first present the observational data and then discuss their interpretation.

\subsection{semi-major axis distribution of stellar companions}

\begin{figure}[htb]
    \centering
    \includegraphics[width=8.5cm]{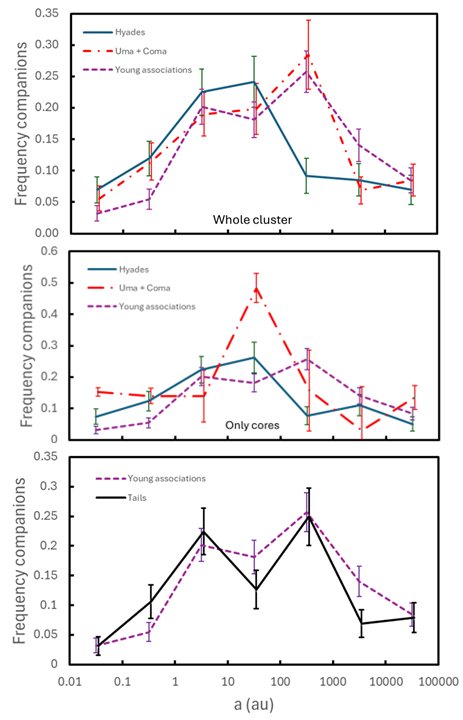}
    \caption{Frequency of companions with semi-major axis $a$ in one-decade bins. The observed frequencies were corrected for incompleteness using the results obtained in Section~\ref{sec:completeness} (averaged over mass ratios $q>0.05$). Upper panel is for the whole samples considered in this paper; middle panel is only for the cores of the different stellar systems; lower panel is for the tails of of the clusters. Blue solid line is for Hyades, red dash-dotted line is for the sum of Ursa Major and Coma Berenices, black solid line is the average of the tails of the clusters, violet short-dashed line is for NYMGs \citep{Gratton2024}. Some of the distributions have been slightly shifted horizontally to allow the appropriate error bars to be distinguished. }
    \label{fig:a_dist}
\end{figure}

\subsubsection{Observational data}

Figure \ref{fig:detection_vs_completeness} shows the distribution of the companions in $a$ (in au) in bins of one-decade for the three clusters, separately. Here we averaged the results obtained for mass ratios $q>0.05$ that are in the stellar mass regime. Companions for which the probability that they might be unrelated stars in the cluster is higher than 50\% were not plotted. The observed values were corrected for incompleteness considering the results of Section~\ref{sec:completeness}. We did this by dividing the observed numbers by the average detection efficiency in the same separation bin; this was only computed for mass ratios $q>0.05$.

The semi-major axis distribution of companions is quite different for Hyades and the remaining clusters considered in this paper. This is clear from the upper panel of Figure \ref{fig:a_dist} that shows the frequency of companions with $a$ in bins of one decade. We found that the frequency of companions at separation $>100$ au corrected for incompleteness is $0.260\pm 0.041$\ for Hyades, while it is $0.495\pm 0.071$\ for the two other clusters. The distribution of the semi-major axis of the companions in the Ursa Major and Coma Berenices samples is not much different from that observed in NYMGs \citep{Gratton2024}; in this last case, the fraction of companions (corrected for completeness) at separation $>100$ au is $0.389\pm 0.037$.

We also note that the fraction of stars still bound to the systems is different in the three samples considered here. If we consider the tidal radius (9 pc, \citealt{Roeser2011}) as the distance from the centre cluster that divides members of Hyades core from those in tidal tails, we found that 129 of the 169 systems with primary mass $>0.8$ \MSun are in the core. This represents 76\% of the stars. On the other hand, only 20\% and 35\% of the stars we considered for Ursa Major and Coma Berenices are in the cores, the others being on the tails of these clusters\footnote{As mentioned in Section~\ref{sec:masses}, only a fraction of the stars in the tails of the three clusters is included in our survey, both because a large fraction of them is at large distances from the Sun at present and the search for lost members is incomplete even at distances from the Sun lower than 100 pc.}. We then plotted the semi-major axis distribution of companions only for stars in the cores of the different stellar systems in the middle panel of Figure \ref{fig:a_dist}. Although the scatter due to small number statistics is quite large, this figure shows that the semi-major axis distribution of companions for the Ursa Major and Coma Berenices core samples (with a median value of 24 au) is intermediate between that observed in Hyades (median value of 14 au) and that obtained for the NYMGs (median value of 108 au)\footnote{We consider the median here because the effect is not very clear and we the prefer to use a more robust statistics.}. On the other hand, the lower panel of Figure \ref{fig:a_dist} compares the semi-major axis distribution of companions to stars in the tails of the programme cluster with that obtained for stars in NYMGs by \citet{Gratton2024}. The two distributions are quite similar, with a small trend for the binaries in the tails of the open clusters (median value of the semi-major axis equal to 52 au) to be more compact than those observed in NYMGs.

\subsubsection{Interpretation}

Differences in the separation distribution of companions for binaries in various environments might be generated at the stellar birth, but they are likely heavily affected by dynamical interactions with other cluster members after that. In fact, as explained by \citet{Heggie1975}, binaries that have low binding energies relative to the kinetic energies of stars within the cluster, known as 'soft' binaries, tend to become even less bound ('softer') on average as a result of interactions with other stars and are eventually disrupted. Binaries with high binding energies relative to the kinetic energies of stars within the cluster ('hard' binaries) conversely become more tightly bound ('harder'). Most hard binaries are unlikely to be disrupted by stellar encounters, but dynamical hardening and binary evolution processes can also destroy very hard binaries. We expect that close encounters occur more frequently in the denser core of the cluster. Offsetting these disruption processes is the preferential tidal stripping of low-mass single stars from the cluster by the Galactic potential and their ejection from the core due to dynamical encounters, which, after an initial stage of rapid soft binary disruption, can lead to a roughly constant global binary frequency over many Gyr \citep{Hurley2005, Geller2013}.

For what concerns binary disruption, the stellar encounter rate with a distance of $d$ in a star cluster is approximately written as the inverse of the time interval between encounters ($t_{\rm enc}$):
\begin{multline}
\frac{1}{t_{\rm enc}}\sim 1.7\times 10^{-6} \left( \frac{n}{10^2\,{\rm star\, pc}^{-3}} \right) \, \left( \frac{v}{1\,{\rm km\,s}^{-1}} \right) \,  \left(\frac{d}{10\,{\rm AU}} \right)^2 \\
\times \left[ 1+89 \left( \frac{M}{1\,{\rm M}_\odot} \right) \left( \frac{v}{1\,{\rm km\,s}^{-1}} \right)^{-2} \left( \frac{d}{10\,{\rm AU}} \right)^{-1} \right] \, {\rm Myr}^{-1}
\end{multline}
\citep{Fujii2019}, where $n$ is the stellar density, $v$ is the velocity dispersion, and $M$ is the mass of a star. If we insert values appropriate for Hyades\footnote{We considered a density of 2.21 stars/pc$^3$ and a velocity spread of 0.45 \kms (see \citealt{Roeser2011}).} into this equation, we find that for a typical stellar system considered here (average total mass of 1.89 \MSun) the time interval between encounters is of the order of the cluster age for an encounter parameter of $d\sim 1015$ au. If we now make the same estimate but using values appropriate for the Coma Berenices cluster\footnote{We considered a density of 0.14 stars/pc$^3$, using radius and mass from \citet{Hunt2024}, and a velocity spread of 0.85 \kms as obtained using Gaia DR3 data for the stars in the core of the cluster considered here.}, the time interval between encounters is of the order of the age of the cluster for an encounter parameter of $d\sim 8700$ au. Considering the rough approximations done, these two values agree reasonably well with the observed truncation of the semi-major axis distribution of companions.

\subsection{Mass ratio distribution of stellar companions}

\begin{figure}[htb]
    \centering
    \includegraphics[width=8.5cm]{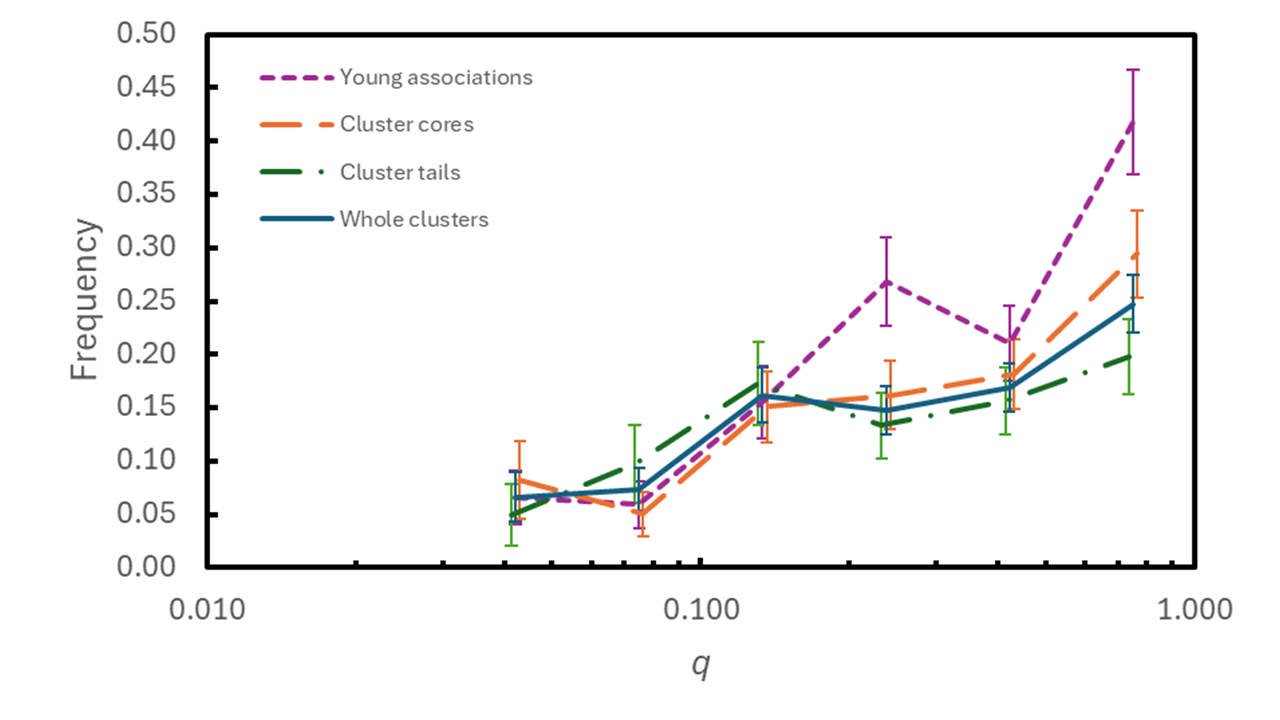}
    \caption{Mass ratio $q$ distribution of the frequency of companions. The observed frequencies were corrected for incompleteness using the results obtained in Section~\ref{sec:completeness} (averaged over semi-major axis $0.01<a<10000$ au). Blue solid line is for combination of all clusters; orange long dashed line is for the cluster cores, green dot-dashed line is for cluster tails, and the violet short-dashed line is for NYMGs \citep{Gratton2024}. Some of the distributions have been slightly shifted horizontally to allow the appropriate error bars to be distinguished.}
    \label{fig:q_dist}
\end{figure}

\subsubsection{Observational data}

Figure \ref{fig:q_dist} shows the mass ratio $q$ distribution of stellar and BD companions (in steps of 0.2 dex in the logarithm of this quantity). In the preparation of this figure, we combined data for the three clusters to reduce the impact of noise, but we considered results obtained for members of the cores and tails apart. We did not plot the points relative to companions for which the probability that they might be unrelated stars in the cluster is higher than 50\%. The observed values were corrected for incompleteness considering the results of Section~\ref{sec:completeness}. We did this by dividing the observed counts by the average detection efficiency in the same separation bin; this was only computed for mass ratios $q>0.05$.

We also show in Figure \ref{fig:q_dist} an analogous distribution obtained for the young association discussed in \citet{Gratton2024}. The observed frequencies were corrected for incompleteness using the results obtained in Section~\ref{sec:completeness} (averaged over $0.01<a<10000$ au). In all cases, the distributions peak towards higher values of $q$, that is, equal-mass binaries are favoured with respect to companions much smaller than the primaries. 

\subsubsection{Interpretation}

The deficiency of equal-mass binaries in the tails with respect to the core is only marginally significant. If real, this small difference might be explained because equal mass binaries are on average systems that are massive and they then more rarely evaporate from a cluster.

More surprising is the fact that the mass ratio distribution of companions for NYMGs is more peaked towards equal-mass members than that obtained for cluster core members. This cannot be explained by the preferential destruction of wide binaries because the expectation is that the mass ratio $q$ decreases with separation. The reduction in the fraction of systems of equal mass may be an effect of companion exchange that is expected for binaries in highly evolved systems \citep{Hills1977, Malmberg2007}. Alternatively, we suggest that this difference might instead be primordial and that binaries born in high-density environments might have on average a lower mass ratio $q$ than those born in low-density environments. This could be explained by three different mechanisms. First, in high-density environments the proto-stellar discs are likely to be destroyed more rapidly (e.g. \citealt{Otter2021}), leaving less time for the secondary to preferentially accrete mass with respect to the primary \citep{Clarke2012, Kratter2010}, which overall results in a lower $q$ value. Since in this case we consider massive companions likely formed by disk instability in very early phases \citep{Vorobyov2006,Machida2010}, the associated disk lifetimes should be very short, likely less than 1 Myr. Alternatively, late events of mass accretion from the interstellar medium onto the disk more easily occur with a different orientation of the angular momentum in more violent environments; in this scenario, we no longer expect that accretion on the secondary is favoured with respect to accretion on the primary \citep{Ceppi2022}, and there is no trend towards equal-mass binaries. Finally, a third possibility is that tidal capture or three-body encounters \citep{Offner2010, Offner2016} more easily occur in dense environments, producing a higher fraction of low-$q$ binaries. However, since the frequency of low-$q$ binaries is similar in the cluster considered in this paper and in the NYMGs (see Figure \ref{fig:q_dist}), this last mechanism should be offset by a general lack of binaries.

\subsection{Overall binary frequency}

\begin{figure}[htb]
    \centering
    \includegraphics[width=8.5cm]{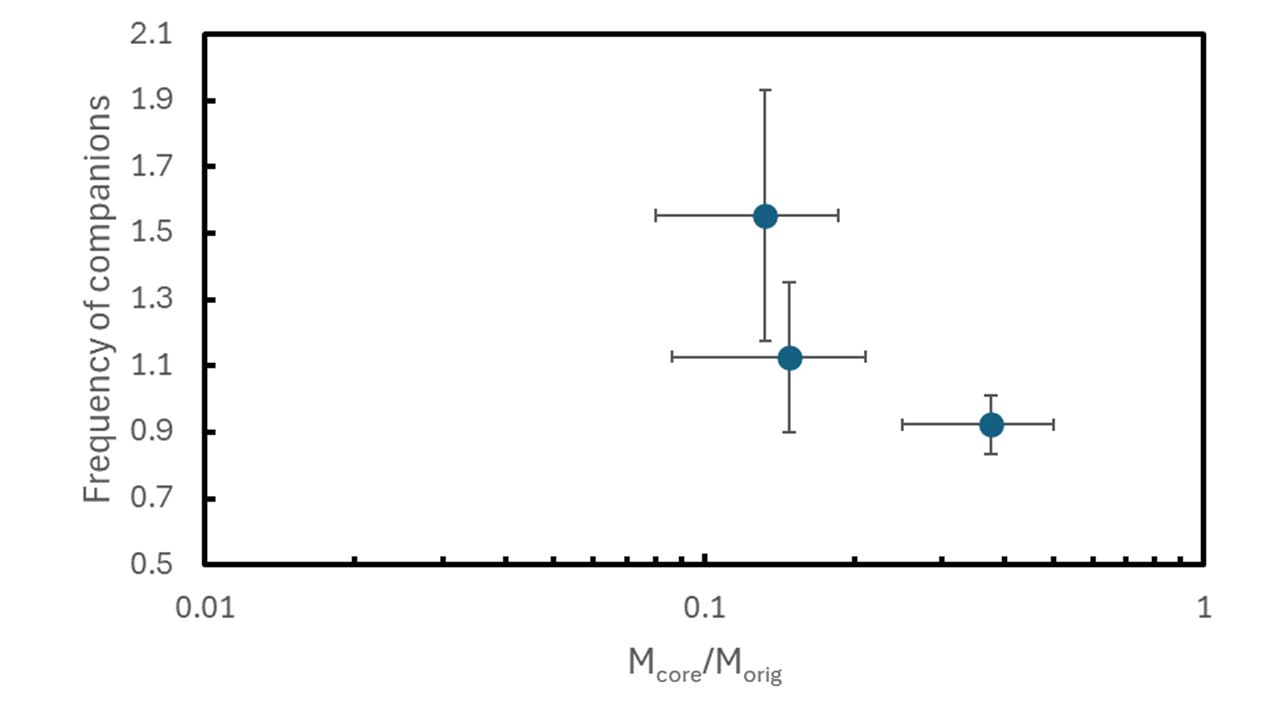}
    \caption{Frequency of companions in the cores of the different stellar systems as a function of ratio between the current mass of the core and the original mass of the cluster.}
    \label{fig:freq_comp}
\end{figure}

\begin{table}
\caption{Frequency of companions (Comp.) for nearby open clusters.}
\begin{tabular}{lccccl}
\hline
Cluster & Primaries & Comp. & Frequency \\
\hline
Ursa Major     &  64 & 56 & $1.05\pm 0.14$ \\
Ursa Major core&  13 & 17 & $1.55\pm 0.38$ \\
Ursa Major tail&  51 & 39 & $0.93\pm 0.15$ \\
\hline
Coma Berenices     & 119 & 68 & $0.96\pm 0.12$ \\
Coma Berenices core& 37 & 25 & $1.13\pm 0.22$ \\
Coma Berenices tail& 82 & 44 & $0.89\pm 0.13$ \\
\hline
Hyades         & 169 & 133 & $0.90\pm 0.08$ \\
Hyades core    & 129 & 104 & $0.92\pm 0.09$ \\
Hyades tail   & 40 &  29 & $0.84\pm 0.16$ \\
\hline
NYMGs& 280 &  221 & $0.98\pm 0.07$ \\
\hline
\end{tabular}
\tablefoot{We reported the observed number of primaries and companions in the second and third column. Companions for which the probability that it might be an unrelated star in the cluster is higher than 50\% were not considered. The frequencies have been corrected for incompleteness using the results of Section~\ref{sec:completeness}  (averaged over mass ratios $q>0.05$). The last row reports the overall result for NYMGs from \cite{Gratton2024}.}
\label{tab:frequency}
\end{table}

\subsubsection{Observational data}

Table \ref{tab:frequency} gives the observed frequency of companions in the three clusters. Companions for which the probability that they might be unrelated stars in the cluster is higher than 50\% were not considered. The observed counts were corrected for incompleteness considering the results of Section~\ref{sec:completeness}. We did this by dividing the observed counts in bins of 0.1 dex in both logarithm of separation and mass ratio, and then dividing for the average detection efficiency computed in the same bin. The corrected frequency is the sum over all bins with mass ratios $q>0.05$ because we are considering here the frequency of stellar companions. Uncertainties in this table are those given by the Poisson statistics of the actual counts and do not include the contribution due to uncertainties in the completeness fractions.

When discussing these frequencies, we should consider the different fraction of objects in the core and tails of the different clusters. We then also give the values obtained only for the cores and tails of the clusters. The frequency of companions in the tails of the clusters is similar to that observed in NYMGs, close to one companion per star. We note that the frequencies of companions in both NYMGs and the tails of the clusters are higher than the value of 0.61 found by \citet{Raghavan2010}. This might be due to a higher degree of completeness in our survey, thanks to the progress in the observational databases cited in the Introduction.

Additionally, as shown in Figure \ref{fig:freq_comp}, the frequency of companions in the cores of the different stellar systems is anti--correlated with the ratio between the current mass of the core and the original mass of the cluster, which is a measure of the dynamical evolution of the system. This frequency is high for Ursa Major and Coma Berenices. 

\subsubsection{Interpretation}

The similar frequency of companions in the tails of the clusters and in NYMGs is expected because dynamical simulations show that the binary fraction beyond the half-light ratio remains constant in clusters \citep{Hurley2007}.

The high binary frequency observed in the cores of systems close to dissolution, such as Ursa Major and Coma Berenices, agrees with expectations for dynamically evolved clusters \citep{Hurley2007}. On the other hand, the frequency of companions in Hyades core is similar to that in the tails, in agreement with expectation for less evolved systems. For a more exhaustive discussion of Hyades dynamics and the related issues, see \citet{Evans2022}.

\begin{table*}[htb]
\centering
\caption{Frequency of JL planets. }
\begin{tabular}{lcccccccc}
\hline
Cluster	& Stars PMa	&Stable	& Fraction	& Detected	&	Dyn	&	Total	&$N_{\rm eff}$ & Freq.		\\
	&		&		&	stable	&	JL comp	&		&	JL comp	&	&	\\
\hline
Ursa Major&	54	&	30	&	0.69	&		&	1	&	1	 & 12.51	&	0.08$\pm$0.08	\\
Coma Berenices	&	52	&	36	&	0.48	&		&	1	&	1	 &	6.81	&	0.15$\pm$0.14	\\
Hyades	& 143	&	68	&	0.69	&	1	&	3	&	4	 & 20.68	&	0.19$\pm$0.09	\\
\hline
\end{tabular}
\tablefoot{First column is the cluster name. Second column the number of stars having the PMa value \citep{Kervella2022}. Third columns give those stars with PMa value for which the orbit of a JL planet can be stable. Column 4 gives the fraction of the stars with potentially stable orbit of JL planets. Column 5 gives the number of planets detected in HCI. Column 6 the number of planets detected though the PMa value. Column 7 the total number of planets to be considered for derivation of the planet frequency. Column 8 gives $N_{\rm eff}$ that is the effective number of stars for which JL planets would be detectable. Column 9 gives the fraction of stars with JL planets.}
\label{tab:jl_frequency}
\end{table*}

\section{Frequency of Jupiter-like planets}
\label{sec:discussion_jl}

\begin{figure}[ht]
\centering
\includegraphics[width=\linewidth]{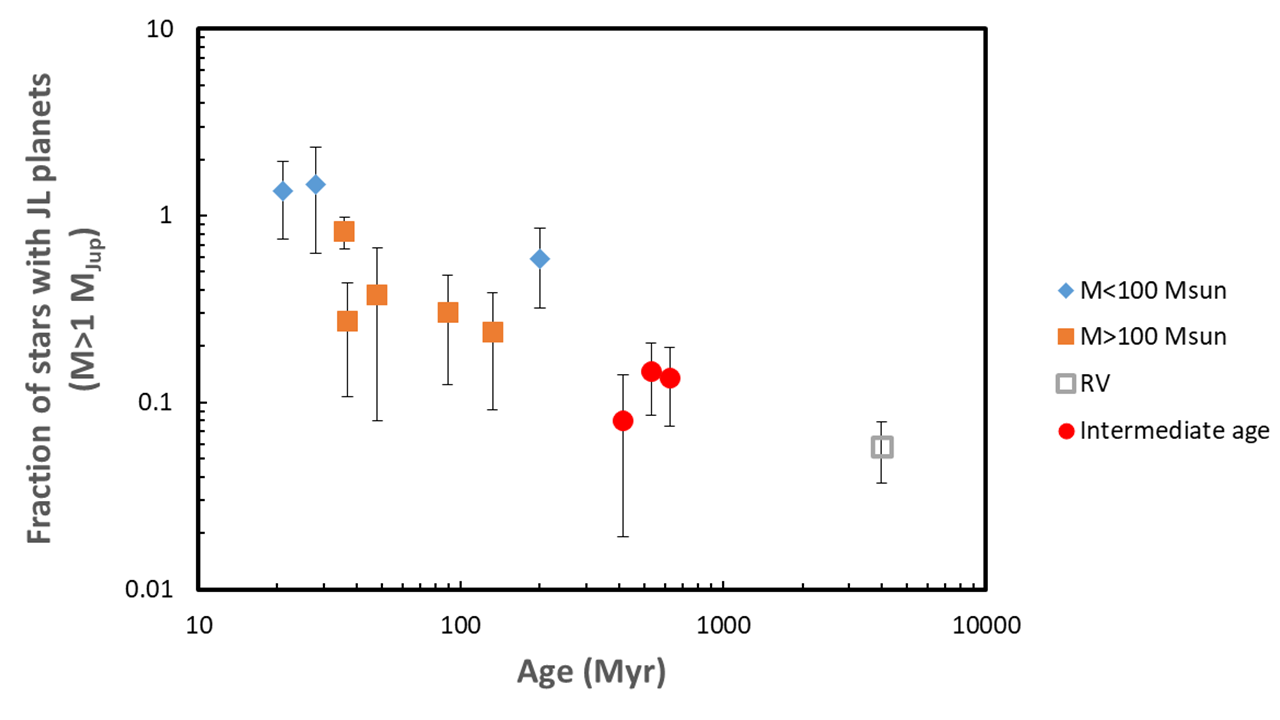}
\caption{Frequency of stars hosting JL planets (corrected for completeness) in individual associations as a function of age. Blue dots are associations with a total mass $<100$ \MSun \citep{Gratton2024}; orange squares are associations with a total mass $>100$ \MSun \citep{Gratton2024}; red circles are the open clusters discussed in this paper. We corrected the observed frequency in Hyades for the higher than solar metallicity of the cluster. The open square represents the frequency of stars hosting JL planets from the RV surveys (adapted from \citealt{Fernandes2019}, see Appendix \ref{appendix:rv_frequency}). We arbitrarily assumed an age of 4 Gyr as typical for the stars in these surveys.
}
\label{fig:frequency_jl_age}
\end{figure}

\begin{figure}[htb]
\centering
\includegraphics[width=\linewidth]{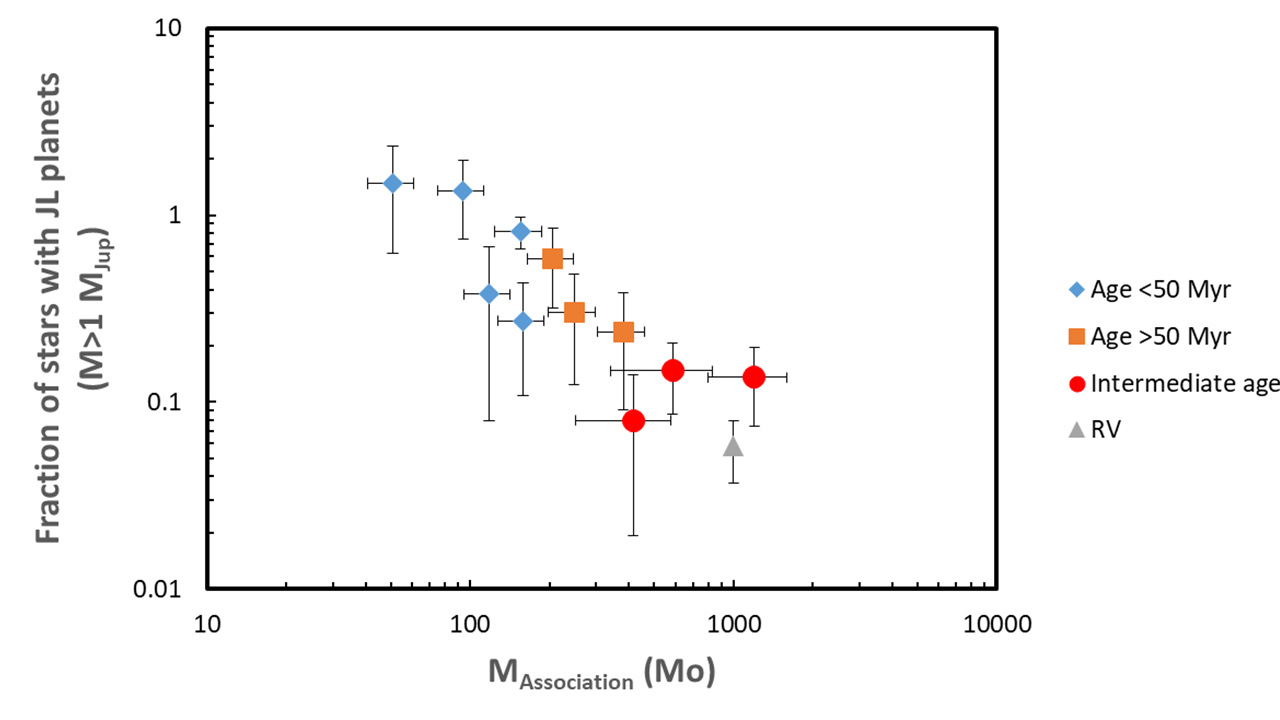}
\caption{Frequency of stars hosting JL planets (corrected for completeness) in individual associations as a function of association mass. Blue dots are associations with an age $<50$ Myr \citep{Gratton2024}; orange squares are associations with a total mass $>50$ Myr \citep{Gratton2024}; red circles are the open clusters discussed in this paper. We corrected the observed frequency in Hyades for the higher than solar metallicity of the cluster. The grey triangle represents the frequency of stars hosting JL planets from the RV surveys (adapted from \citealt{Fernandes2019}, see Appendix \ref{appendix:rv_frequency}). We arbitrarily assumed a mass of 1000 \MSun for the typical birthplace of stars in these surveys.
}
\label{fig:frequency_jl_mass}
\end{figure}

In this Section we discuss our results concerning JL planets. The semi-major axis and mass ratio distributions of binaries in the intermediate-age open clusters studied here show properties similar to those in NYMGs for stellar companions within 100 au, despite the complexity related to dynamical evolution. We might then expect that this might also occur for the JL planets since their semi-major axis is below this limit. This also agrees with the numerical simulations of \citet{Fujii2019} where $<1.5$\% of close-in planets within 1 AU and at most 7\% of planets with 1-10 AU are ejected by stellar encounters in clustered environments. It may then be interesting to compare their frequency with that obtained for NYMGs.

Given the old ages of the populations considered in this paper, JL planets can be discovered using dynamical methods, either from high-precision RV or PMa (see Section~\ref{sec:substellar_companions}), while they are very difficult to observe using HCI (see, however, the case of HIP 21152B: \citealt{Bonavita2022, Franson2023c}). However, high-precision RVs are available for less than 50\% of the stars, and in many cases the time series are not long enough to extensively cover the semi-major axis range appropriate for JL planets. This makes completeness corrections more uncertain. We then mainly focus on results obtained using PMa, which are available for about 70\% of the stars and is sensitive to systems with semi-major axis that rather well match the range of JL planets. This is also appropriate in the comparison with NYMGs because PMa data were available for a similar fraction of stars in that case, and they are unbiased with respect to age (at variance with RV).

\subsection{Methods used to estimate the frequency of Jupiter-like planets}

In this Section we present the method used to estimate the frequency of stars with at least one JL planet in the intermediate age clusters. The method used is identical to that used to estimate the frequency of JL planets in NYMGs in \citet{Gratton2024}, but we use a somewhat extensive definition of the mass and separation range for JL planets ($1.5<a<50$, $0.001<q<0.02$). Note that there are no additional JL planetary systems in NYMGs when using this definition; there is then no loss of consistency due to this assumption. A total of six objects satisfy these conditions in the three clusters; they all have the PMa measured. This should then be compared with the total number of stars (249) for which a PMa measure has been obtained. 

We should also consider that for more than half of these stars, there are stellar companions that would make the orbit of JL planets unstable and hence cannot have a JL planet. We estimated for this purpose the inner and outer radii of the instability regions around these companions using the formulation of \citet{Holman1999}. We assumed here an orbit eccentricity given by equation (1) in \citet{Gratton2023b}. Driven by the results obtained this way, as in \citet{Gratton2023b, Gratton2024} we assumed that the orbit of a JL planet would be unstable around any star for which an inner companion would cause $a_{\rm max}>1.5$ au or an outer companion $a_{\rm min}<20$ au. Here, $a_{\rm min}$ and $a_{\rm max}$ are the inner and outer edges of the region where orbits are unstable due to the perturbations by a companion\footnote{Strictly speaking, when we adopt a value of $a_{\rm min}<20$ au there may be cases where JL planets with $20<a<50$ au are on unstable orbits. On the other hand, there may be stable orbits for JL planets according to the criteria by \citet{Holman1999} even when these conditions are not met over the whole range considered here. However, application of more precise criteria would require knowledge of the distributions of JL planet orbits as a function of the semi-major axis and eccentricity that are likely affected by the existence of stellar companions \citep{Kozai1962, Rasio1996, Ford2000}. In any case, these orbits would be close to the edge of the stability region and we preferred to use the same criteria considered in \citet{Gratton2023b, Gratton2024} for consistency.}. We note that the orbital stability radius typically depends on the 1/3 power of the mass and linearly on the orbital semi-major axis of the companion. Since the semi-major axis distribution peaks at a few tens au, the companions responsible for instability of a JL planet are on wide orbit in most cases and then the relevant objects are mainly visual binaries with a small contribution by astrometric binaries. We give estimates of typical errors for the masses and semi-major axis derived with different methods in Section \ref{sec:companion_parameters}. We found typical errors of about 40\% in the semi-major axis for both visual and astrometric binaries; errors in the masses are about 40\% of the astrometric binaries but they are negligible for the visual binaries. We may then assume that the main source of uncertainty in the estimate of stability radii is the semi-major axis of the companion orbit. We found that there are 114 stars for which the orbit of a putative JL planet would be unstable with the stability limits determined this way and that this count is uncertain by less than 5\% if errors are randomly distributed.

Consideration of orbital stability leaves a total of 134 stars that have a PMa value and potentially host a JL planet. Furthermore, most remaining stars are at quite large distances from the Sun, so that only massive JL planets can be detected using PMa. As discussed in \citet{Gratton2024}, the fraction of planets that can be detected in this way depends on the limiting mass $M_{min,i}$ for this technique, which in turn depends on many factors, mainly distance from the Sun. According to our assumption, a companion is detected when the star has $SNR(PMa)>3$. \citet{Kervella2022} lists the $SNR(PMa)$ value for all stars, as well as the masses corresponding to the observed signal at 3, 5, 10 and 30 au, in their Table A.1. Let us call $M(a)$ these values. We may then assume that the smaller companion detectable with PMa at a separation $a$ using PMa has a mass equal to $M_{min}=(3/SNR) M(a)$ as listed in their table. Since most JL planets have semi-major axis in the 5-10 au range (see Appendix~\ref{appendix:rv_frequency}), we adopted $M_{min}=(3/SNR) [M(5)+M(10)]/2$. We then considered the fraction of JL planets around each star that are expected to be above this limiting mass. This depends on the assumed mass function $f(m)$ for JL planets. As in \citet{Gratton2023b, Gratton2024}, we adopted a mass function that is a power law with a slope of -1.3 as proposed by \citet{Adams2021}, We then call $N_{\rm eff}$ the quantity:
\begin{equation}
N_{\rm eff} = \sum{\frac{\int_{M_{min,i}}^{20}f(m)dm}{\int_{1}^{20}f(m)dm}},
\end{equation}
where the index $i$ refers to each star and $m$ is in units of \MJup. $N_{\rm eff}$ represents the effective number of stars around which JL planets could be found using the PMa technique for each association. The frequency of JL planets is then given by the ratio between the number of actual discoveries and $N_{\rm eff}$. 

We summarised the relevant data in Table \ref{tab:jl_frequency}. With these assumptions, combining the three clusters, we obtain $N_{\rm eff}=40.00$ and an overall frequency of $0.15\pm 0.05$. For the three individual clusters we obtain frequencies of $0.08\pm 0.08$, $0.15\pm 0.14$, and $0.19\pm 0.09$ for Ursa Major, Coma Berenices, and Hyades, respectively. The higher value obtained for Hyades may at least in part be due to their higher metallicity. In fact, using the metallicity dependence of the frequency of giant planets obtained by \citet{Johnson2010}, we expect that this is a factor of 1.43 higher for Hyades metallicity of [Fe/H]=0.13 (see Section~\ref{sec:metallicity}) than it would be for a solar metallicity value. This may introduce a bias in our discussion, where the metallicity dependence is not considered and populations that have the same metallicity should be compared with each other. Since all other clusters considered in this paper and in \citet{Gratton2024} have solar metallicity, we corrected the JL planet frequency of Hyades down to $(0.19\pm 0.09)/1.43=0.13\pm 0.06$ for this factor in the comparison between frequency of JL planets in different associations. Once this effect is considered, the weighted average of the three clusters is $0.12\pm 0.05$. Although the definition of JL planets adopted here is looser than that considered in \citet{Gratton2023b, Gratton2024}, these values are lower than those obtained for NYMGs. They are similar to those obtained from RV surveys \citep{Cumming2008, Mayor2011, Wittenmyer2016, Fernandes2019, Zhu2022, Wolthoff2022}. In fact, as discussed in Appendix~\ref{appendix:rv_frequency}, the results of these surveys indicate that $5.8\pm 2.2$\% of Sun-like stars host JL planets that have $0.001<q<0.02$ and $1.5<a<50$ au. 

\subsection{Results: Trends with mass and age}

The results of the previous subsection confirm then a general trend for decreasing frequency of JL planets with increasing age and or mass of the original populations. This is shown in Figure \ref{fig:frequency_jl_age} and Figure \ref{fig:frequency_jl_mass} for age and mass, respectively. Various possible explanations for these trends are discussed in \citet{Gratton2023b, Gratton2024}. These trends can be represented by the following regression equations:
\begin{equation} \label{eq:frequency_w_mass}
\log{f(M_{\rm orig})} = (0.01\pm 0.11)-( 0.94\pm 0.21) (\log{M_{\rm orig}}-2.0)
\end{equation}
for the original mass $M_{\rm orig}$ in units of \MSun, with a normalised $\chi^2=0.45$, and:
\begin{equation} \label{eq:frequency_w_age}
\log{g(t)} = -(0.36\pm 0.08)-( 0.67\pm 0.14) (\log{t}-2.0) .
\end{equation}
with the age $t$ in units of Myr, with a normalised $\chi^2=0.57$. We note that a trend with mass would likely be mostly related to the formation phase, while that on age would likely be a signature of dynamical effects active on a long time scale (several hundred millions of years) that are expected in open clusters. Numerical simulations by \citet{Fujii2019} found that only a small fraction of planetary systems having a separation below 10 au are expected to be destroyed in a cluster similar to Hyades. However, this depends on the initial density that is not well known and refers to single planets. We note that both the trends with mass and age are highly significant, well above $3-\sigma$. Unfortunately, there is also a high correlation between ages and masses for the clusters/associations considered so far: the linear Persson coefficient is $r=0.926$ if we use the logarithm of the two quantities, and this makes it difficult to conclude which of the two is most relevant using these data alone. Despite this, we might try to separate the dependence on these two quantities; for this we considered the combined regression on both mass and age, that is,
\begin{multline}
\log{h(m,t)} = (-0.10\pm 0.20)-( 0.63\pm 0.45) (\log{M_{\rm orig}}-2.0)-\\
( 0.23\pm 0.33) (\log{t}-2.0) ,
\end{multline}
with normalised $\chi^2=0.49$. However, given the very large error bars and the fact that the normalised $\chi^2$ is not lower than for the regression on the individual quantities, this last result is not very useful and the ambiguity cannot be solved.

\subsection{Discussion of results for Jupiter-like planets}

\subsubsection{Comparison with the frequency of Jupiter-like planets from RV surveys}

These relations allowed us to check if the observed frequency of JL planets in NYMGs and  nearby open clusters of intermediate age is compatible with that obtained from the RV surveys, which for our definition of JL planets is $5.8\pm 2.2$\% as discussed in the Appendix~\ref{appendix:rv_frequency}. If we assume that the frequency of JL planets $f(M_{\rm orig})$ only depends on the initial mass of the cluster or association, the expected frequency of planets from RV surveys $\nu$ should be related to that obtained from Eq.~\ref{eq:frequency_w_mass} through the relation:
\begin{equation}
\nu={\int_{m_{min}}^{m_{max}} f(M_{\rm orig})\, M_{\rm orig}\, \phi(M_{\rm orig}) \, dm}  ,
\end{equation}
where $f(M_{\rm orig})$ is the frequency of JL planets as a function of original mass of the association/cluster $m_{\rm orig}$ (see Eq.~\ref{eq:frequency_w_mass}) and $M_{\rm orig} \phi(M_{\rm orig})$ is the normalised cluster initial mass function. On the other hand, if we assume that the frequency of JL planets $\nu (M_{\rm t})$ only depends on the age of the cluster/association, the expected frequency of planets from RV surveys $\nu$ should be related to that obtained from Eq.~\ref{eq:frequency_w_age} through the relation:
\begin{equation}
\nu={\int_{t_{min}}^{t_{max}}g(t)\, \psi(t)\, \rho(t)\, dt} ,
\end{equation}
where $g(t)$ is the frequency of JL planets as a function age of the association/cluster (see Eq.~\ref{eq:frequency_w_age}), $\psi(t)$ is the normalised star formation rate in the disk of the Milky Way (starting from present towards the past), and $\rho(t)$ is a function that takes into account the possible selection effects against detection of young JL planets from RV surveys due to the higher value of the jitter caused by stellar activity (e.g. \citealt{Queloz2001, Lagrange2023}). 

The logarithmic embedded cluster initial mass function $M_{\rm orig}\, \phi(m_{\rm orig})$ is often thought to be flat for masses $50<M_{orig}<3 \cdot 10^5$ \MSun \citep{Hunter2003,Lada2003}\footnote{This means that clusters within a given interval of mass in logarithm contribute evenly to the total number of stars.}. This corresponds to an embedded cluster mass function with a spectral index of $-2$ (i.e., $dN/dM \approx M^{-2}$). If we consider the dependence of the frequency of JL planets on the mass given by Eq.~\ref{eq:frequency_w_mass} and use a flat initial cluster mass function over the mass range defined above, we find that the expected overall frequency of JL planets in the field population should be $21.1\pm 4.8$\%. The result is slightly lower ($19.0\pm 4.0$\%) if we use the embedded cluster mass function by \citet{Lada2003}, which slightly penalises low-mass clusters. These frequencies of JL planets are larger than that given by the RV survey; this is because according to Eq.~\ref{eq:frequency_w_mass} so many JL planets formed in the lowest mass groups that the weighted average over the whole range of masses would still exceed the frequency deduced from RVs. A very low frequency ($2.1\pm 1.8$\%) would rather be obtained using the initial cluster mass function by \citet{Piskunov2008}. However, this is not appropriate here because it describes the cluster mass function after the infant mortality phase, that is, after the expulsion of gas that causes the preferential dissolution of low-mass clusters, as discussed by the authors. These authors proposed that the high mass slope of $dN/dM\approx M^{-1.66}$ should be more representative of the initial distribution of star-forming regions; this slope is similar to that observed for giant molecular clouds \citep{Blitz2006}. If we use this slope and assume that star-forming regions have masses in the range $50<M<3 \cdot 10^5$ \MSun (as proposed by \citealt{Piskunov2008}), the expected overall frequency of JL planets would be $6.2\pm 2.8$\%, which is in agreement with the result from RV surveys within the errors. We note that if this interpretation is correct, 50\% of the JL planets formed in environments with original stellar mass less than 150 \MSun, 70\% in those with a mass less than 300 \MSun (a typical NYMG)  and only 10\% in those with more than 2,000 \MSun, which might be the birthplace of the Sun \citep{Pfalzner2013,Pfalzner2020,PortegiesZwart2019}.

On the other hand, if we assume that the RV survey targets are distributed according to the star formation history derived using Gaia data \citep{Alzate2021}, neglect any selection effect against detection of young planets, and use the dependence of planet frequency on age given by Eq.~\ref{eq:frequency_w_age}, we should expect that $4.0\pm 2.3$\% of solar-type stars host JL planets. This value is lower than, but consistent with that obtained from the RV surveys. It should be further reduced if we assume that $\rho(t)$ is not constant but  increase with age $t$, as it likely is \citep{Lagrange2023}. This discrepancy could be solved by assuming that the dependence of JL planet frequency on age cannot be extrapolated at ages older than that of Hyades and becomes flatter at old ages. This is possibly realistic because we might expect that the dynamical disruption of planetary systems becomes ineffective once clusters are dissolved.

Finally, it is possible that both mass and age are important and that a suitable combination of the various effects can explain the difference observed between the results we obtained for the young and intermediate-age associations and clusters, and what is observed for (older) stars in the RV surveys.

\subsubsection{Jupiter-like and free floating planets}

If most JL planets are lost by planetary systems on a long time scale ($>20$ Myr), we expect that they become Jupiter-like free floating planets (JFFPs), that is, free-floating planets with a mass in the range of 1-20 M$_{\rm Jup}$. On a long time scale ($\sim 1$ Gyr) this mechanism should produce many JFFPs that should add to the population of JFFPs that are formed in isolation or are lost very early in star-forming regions \citep{Miret-Roig2022, Miret-Roig2023}. We further note that JFFPs lost by planetary systems likely have a different mass function from those formed in isolation. If we adopt the Chabrier IMF \citep{Chabrier2003} we find that 11\% of the stars are solar-type stars (that is, in the $0.8<M<2.3$ \MSun  mass range considered in this paper)\footnote{This fraction is 16\% among the members of Upper Scorpius listed by \citet{Miret-Roig2022}.}. If we also use the dependence of the giant planet frequency on mass by \citet{Johnson2010} we find that they host 39\% of the JL planets. If we now assume that at an age of 10 Myr every solar-type star hosts at least a JL planet (as given by Eq.~\ref{eq:frequency_w_age}), and that the frequency of JL planets with host mass scale as given \citet{Johnson2010}, there should be about $0.28$ JL planets per main-sequence star at this early age. Since Eq.~\ref{eq:frequency_w_age} indicates that $>90$\% of the JL planets are lost in less than 1 Gyr, we should expect an excess of $\gtrsim 0.25$ JFFPs per main-sequence star in old populations with respect to young associations.

JFFPs are quite numerous in young associations. An estimate of their frequency can be obtained from the survey in Upper Scorpius by \citet{Miret-Roig2022}. They listed 2725 stellar (mass $>0.075$ \MSun) members of the Upper Scorpius association, assuming an age of 5 Myr. In this association and assuming the same age, they also listed 234 JFFPs with a mass $<20$ M$_{\rm Jup}$, corresponding to our definition of JL planets. However, their sample of JFFPs only includes objects with mass $>6$ M$_{\rm Jup}$. If we assume that all objects above this mass are detected, and all those below are not, and that the distribution of JFFPs with mass follows the same slope as considered by \citet{Sumi2023} for the low-mass tail of the stellar/BD IMF, \citet{Miret-Roig2022} detected about 40\% of the JFFPs. This would yield about $0.22\pm 0.03$ JFFPs per main-sequence star. This result would change by less than 15\% adopting ages of 3 or 10 Myr, a reasonable range for the Upper Scorpius association \citep{Pecaut2012, Squicciarini2021}. Since Upper Scorpius is actually younger than the NYMGs considered in \citet{Gratton2024}, this potentially large population of JFFPs cannot be due to the long-term loss of JL planets that is given by Eq.~\ref{eq:frequency_w_age}. 

The population of old JFFPs may be derived from the frequency of microlensing events that are of short duration (1-2 days). This method cannot distinguish between free-floating or wide-orbit planets; HCI surveys can be used to constrain the contribution by wide-orbit objects \citep{Clanton2017, Poleski2021, Mroz2024}. The frequency of these short events is somewhat controversial. Although \citet{Sumi2011} concluded that there is more than one JFFP per main-sequence star, the more recent analysis by \citet{Mroz2017} with a much wider sample and sensitivity yields a much lower best value of 0.05 and a 95\% per cent confidence upper limit of 0.25 free-floating or wide-orbit massive planet per main-sequence star. Even more recently, \citet{Sumi2023} considered two populations of objects in the JFFP mass range. The first population is the low-mass tail of the stellar/BD distribution. Integrating the IMF they found for this population over the range 1-20 M$_{\rm Jup}$ yields a frequency of $0.26\pm 0.05$ JFFP per main sequence star. This result is consistent with what we obtained above starting from the results by \citet{Miret-Roig2022} for Upper Scorpius. The second population is given by the high-mass tail of the population of the low-mass free-floating planets. Integrating the IMF they propose for this population over the mass range 1-20 M$_{\rm Jup}$ yields a frequency of $0.064\pm 0.016$ JFFP per main-sequence star. This might represent the possible contribution of planets lost on a long time scale. However, we should correct this frequency downward for the possible contribution of wide-orbit objects. In addition, we should consider that objects detected through microlensing are closer to the galactic centre and then likely more metal rich (due to the galactic metal abundance gradient) than the Sun. Since planet frequency is a strong function of the metallicity of the hosting star, there are likely more planets in more metal-rich environments. We should then consider the frequency of $0.064\pm 0.016$ JFFP per main-sequence star cited above as an upper limit for planets lost by planetary systems in the solar neighbourhood. This then represents only a small fraction of the excess of $\gtrsim 0.25$ JFFP per main-sequence star in old populations expected if age is the driving parameter for the relations discussed in Section 6.4.2.

We conclude that the observed trend in the frequency of JL planets in stellar groups in the solar neighbourhood derived in \citet{Gratton2024} and in this paper is probably mainly due to the mass of the associations/clusters where they formed. Given the uncertainties, there is still room for some contributions due to the long-term loss of planets. 

\section{Conclusions}
\label{sec:conclusions}

In recent papers \citep{Gratton2023b, Gratton2024} we showed that stellar and substellar companions for nearby (distance $<100$ pc) young (age $<200$ Myr) associations roughly split into two separate groups: stellar and BD companions, with mass ratio $q>0.05$, and JL planets with $q>0.02$ that mainly have orbital semi-major axis in the range $1.5<a<50$ au. We also found that the frequency of JL planets in young associations is larger than the one previously reported by RV and decreases with the mass and or age of the association. 

In this paper, we derived the frequency of stellar and JL companions around solar-type stars in the nearby intermediate-age clusters Hyades, Coma Berenices, and Ursa Major. These clusters have ages in the range 400-700 Myr and at their origin have larger masses than the NYMGs. They are then more representative of the environments in which most stars form. The methods used are similar to those considered for the NYMGs, although the older age makes HCI less effective. We estimated the completeness of the survey as a function of $a$ and $q$ considering each of the discovery methods used.

We obtained samples of stellar companions for Hyades and Ursa Major group that were fairly complete; the results are somewhat less complete for the Coma Berenices cluster because of its larger distance. As detailed in Section \ref{sec:substellar_companions}, a few massive JL planets were detected. As in the case of the NYMGs, we found a clear separation of the companion population into two groups. We can explain the properties (frequency and distribution with semi-major axis) of the stellar/BD companions by the dynamical evolution within the clusters. We found a relative scarcity of companions of equal mass that might be related to the formation of binaries in these environments; a concentration of binaries in the cores of Ursa Major and Coma Berenices, attributed to selective loss of low-mass systems; and a lack of wide companions ($a>1000$ au) in Hyades that can be attributed to destruction of binaries by close encounters. However, the population of JL planets (with semi-major axis $a<50$ au) should not be heavily affected by this phenomenon. We found that the frequency of JL planets is lower than in the NYMGs and rather similar to that obtained from RV surveys. We derived empirical relations connecting the original mass, age, and frequency of JL planets.

We noted that once the initial mass function of clusters and star formation history in the galactic disk are considered, the frequency of JL planets in NYMGs is consistent with that found from RVs if we assume that a significant fraction of the JL planets are lost in time and or their original frequency depends on the mass of the environment following the empirical relations described above. We note that if the last is the correct interpretation of these data then 50\% of the JL planets formed in environments with original stellar mass less than 150 \MSun, 70\% in those with a mass less than 300 \MSun (a typical NYMG)  and only 10\% in those with more than 2,000 \MSun that might be the birthplace of the Sun \citep{Pfalzner2013,Pfalzner2020,PortegiesZwart2019}.

Unluckily, given the strong correlation existing between mass and age in the stellar groups considered so far, we are unable to remove the degeneracy currently existing between mass and age and correctly weight their relevance using our results alone. In principle, this degeneracy can be removed by exploring either old sparse moving groups or very young open clusters. In a future paper we plan to consider field stars that have ages similar to that of the intermediate-age open clusters considered in this paper and since they are more dispersed possibly mainly formed in lower density environments. However, since their birthplaces are not well defined and the ages bear considerable uncertainties, their use for understanding the impact of the environment on planet formation should be considered carefully. Alternatively, it might be possible to extend the search for JL planets to massive, young systems. Examples of similar populations exist further than 100 pc (e.g. the Pleiades or the Ophiucus star-forming complex), but at such large distances the search of JL planets becomes difficult with current instrumentation. For instance, at the distance of about 140 pc of these stellar groups JL planets at separation of 10 au appear projected at only 70 mas from the star, behind the coronagraphic mask of high contrast imagers on 8-10 m class telescopes. Even the Nancy Grace Roman telescope \citep{Kasdin2020}, which may detect JL planets around nearby stars, has an inner working angle of about 0.15 arcsec, which makes it less useful for studying them further than 100 pc from the Sun. At this distance, a very massive JL planet at the upper edge of the considered distribution ($q=0.02$) is related to a PMa with a $SNR\sim 3.5$. Hence surveys of the Pleiades and Ophiucus would not be able to detect such objects with current instrumentation. Such objects are possibly beyond the limit even of future Gaia data releases that according to current predictions may discover JL planets only out to 50 pc distance \citep{Lattanzi2010, Perryman2014}. Detecting such objects should be rather possible with high contrast imagers on telescopes of the 30-40 meter class such as the E-ELT that can well explore the stellar neighbourhood of young stars down to $\sim 40$ mas or even closer in.

An alternative way to separate the dependence on mass from that on age can be provided by examining the frequency of Jupiter-mass free floating planets (JFFP) in different environments. In fact, if a large fraction of the JL planets is lost on a long-term scale, this mechanism alone should produce an excess of $\gtrsim 0.25$ JFFP per main-sequence star in old populations with respect to young ones. We then compared the frequency of JFFP in young associations (e.g. in Upper Scorpius: \citealt{Miret-Roig2022}), formed on a short time scale, with that in old populations as obtained from microlensing (e.g., \citealt{Mroz2017, Sumi2023}), and checked for this excess. The current evidence is that there is only a limited excess of JFFP in old populations (at most $0.064\pm 0.016$ JFFP per main-sequence star). If confirmed, this comparison suggests that only a few JL planets are lost on a long-term scale and then that the observed trends are mainly due to the mass of the environment where stars form.

Finally, further progress can be obtained by a better understanding of the survival of disks in high-density environments, such as the Orion cluster (e.g., \citealt{Ansdell2017, Ansdell2020, vanTerwisga2019,vanTerwisga2020}). Studies of the dynamical evolution of planetary systems in open clusters (e.g. \citealt{Fujii2019}) could also be very useful. In addition, it would be useful to have a systematic comparison of planet properties in stars of different ages in the RV surveys samples. 

\begin{acknowledgements}
This work has made use of data from the European Space Agency (ESA) mission {\it Gaia} (\url{https://www.cosmos.esa.int/gaia}), processed by the {\it Gaia} Data Processing and Analysis Consortium (DPAC, \url{https://www.cosmos.esa.int/web/gaia/dpac/consortium}). Funding for the DPAC has been provided by national institutions, in particular, the institutions participating in the {\it Gaia} Multilateral Agreement.
This research has used the SIMBAD database, operated at CDS, Strasbourg, France. 
D.M., R.G., and S.D. acknowledge the PRIN-INAF 2019 'Planetary systems at young ages (PLATEA)' and ASI-INAF agreement n.2018-16-HH.0. A.Z. acknowledges support from ANID -- Millennium Science Initiative Program -- Center Code NCN2021\_080. S.M.\ is supported by the Royal Society as a Royal Society University Research  Fellowship (URF-R1-221669). V.S. acknowledges support from the European Research Council (ERC) under the European Union’s Horizon 2020 research and innovation programme (COBREX; grant agreement n◦ 885593). V.D. acknowledges the financial contribution from PRIN MUR 2022 (code 2022YP5ACE) funded by the European Union – Next Generation EU.
SPHERE is an instrument designed and built by a consortium consisting of IPAG (Grenoble, France), MPIA (Heidelberg, Germany), LAM (Marseille, France), LESIA (Paris, France), Laboratoire Lagrange (Nice, France), INAF-Osservatorio di Padova (Italy), Observatoire de Gen\`eve (Switzerland), ETH Zurich (Switzerland), NOVA (Netherlands), ONERA (France) and ASTRON (Netherlands), in collaboration with ESO. SPHERE was funded by ESO, with additional contributions from CNRS (France), MPIA (Germany), INAF (Italy), FINES (Switzerland) and NOVA (Netherlands). 
This research has made use of the Exoplanet Follow-up Observation Program (ExoFOP; DOI: 10.26134/ExoFOP5) website, which is operated by the California Institute of Technology, under contract with the National Aeronautics and Space Administration under the Exoplanet Exploration Program.

\end{acknowledgements}

\bibliographystyle{aa} 
\bibliography{main} 

\begin{appendix}

\section{On the frequency of Jupiter-like planets from RV surveys}
\label{appendix:rv_frequency}

\citet{Fernandes2019} give a frequency of $6.2^{+1.5}_{-1.2}$\% for planets with masses in the range 1-20 M$_{\rm Jup}$ and $0.1<a<100$ au from RV data. This result cannot be directly compared with the frequency of JL planets obtained for NYMGs because the definition of JL planet they used is different from the one used to obtain the values we reported for the NYMGs regarding the semi-major axis ranges 
for which we rather considered the range $1.5<a<50$ au. To estimate the correction due to this different semi-major axis range, we first examined the double symmetric power-law distribution of companions  period with slope of (0.65, -0.65) centred at a period P=1580 d considered by \citet{Fernandes2019} and found that 27\% of the planets they considered are in the semi-major axis $0.1<a<1.5$ au (that is below or lower limit), 59\% in the range $1.5<a<10$ au, 12\% in the range $10<a<50$ au, and 2\% in the range $50<a<100$ au (that is above our higher limit). We then noted that the broken power-laws approach by \citet{Fernandes2019} yields few planets in the $10<a<50$ au range with respect to the $1.5<a<10$ au. However, RV surveys do not really explore this last range of semi-major axis and the corresponding frequency is extrapolated from the results obtained at shorter periods. On the other hand, we may note that \citet{Gratton2024} detected 13 JL planets with $a<10$ au and 16 with $a>10$ au in NYMGs; these results are more appropriate for long period planets. If we correct these counts for completeness over the range $0.001<q<0.02$ appropriate for JL planets (see Figure 10 in \citealt{Gratton2024}) and assuming that JL planets distribute in mass as proposed by \citet{Adams2021}, we obtain that in the sample of NYMGs there should be 53 JL planets with $a<10$ au and 31 with $a>10$ au, that is, about $37\pm 14$\% of the JL planets have $a>10$ au. Since RV may easily miss planets with very long periods, the extrapolation of the power-law approach at long periods may be inaccurate. If we rather assume that the fraction of JL planets with $a<10$ au is the same for the population of planets in NYMGs and RV surveys, we may combine the correction in mass and semi-major axis and conclude that existing data suggests a frequency of $5.8\pm 2.2$\% for JL planets (more massive than 1 M$_{\rm Jup}$) around the stars examined in the RV surveys\footnote{The quoted uncertainty is the quadratic sum of the statistical one inherited from the paper by \citet{Fernandes2019} and of the uncertainty in the correction for the distribution in semi-major axis.}.

\section{Gaia DR3 detection limits}
\label{appendix:dr3_detlim}

\begin{figure}[htb]
\centering
\includegraphics[width=\linewidth]{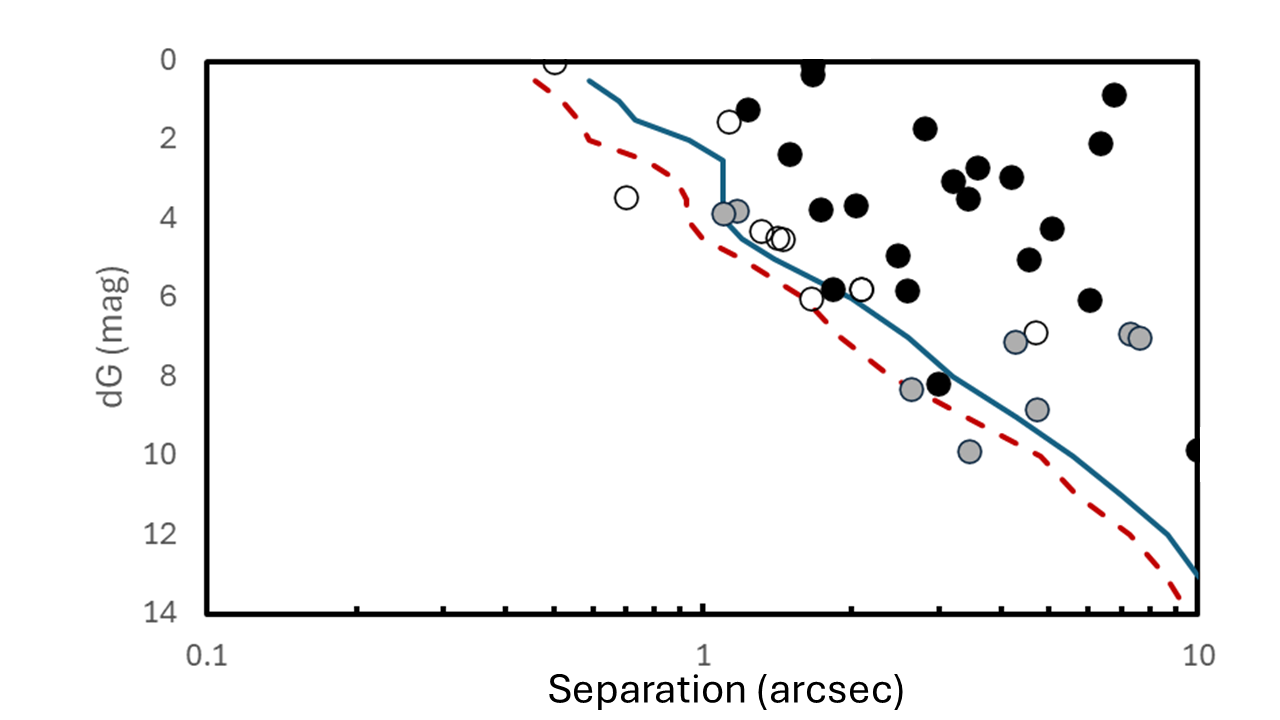}
\caption{Run of the contrast dG versus separation in arcsec for companions detected in Gaia DR3. We plotted only companions with separation $<10$ arcsec. Black filled circles are companions also detected with full astrometric solution in Gaia DR2. Grey filled circles are companions detected in Gaia DR2 but without astrometric solution. Empty circles are companions not detected in Gaia DR2. The blue solid and red dashed lines are the 50\% and 1\% contrast limits for Gaia DR2 according to \citet{Brandeker2019}, respectively.
}
\label{fig:gaia_dr2_dr3}
\end{figure}

\begin{figure}[htb]
\centering
\includegraphics[width=\linewidth]{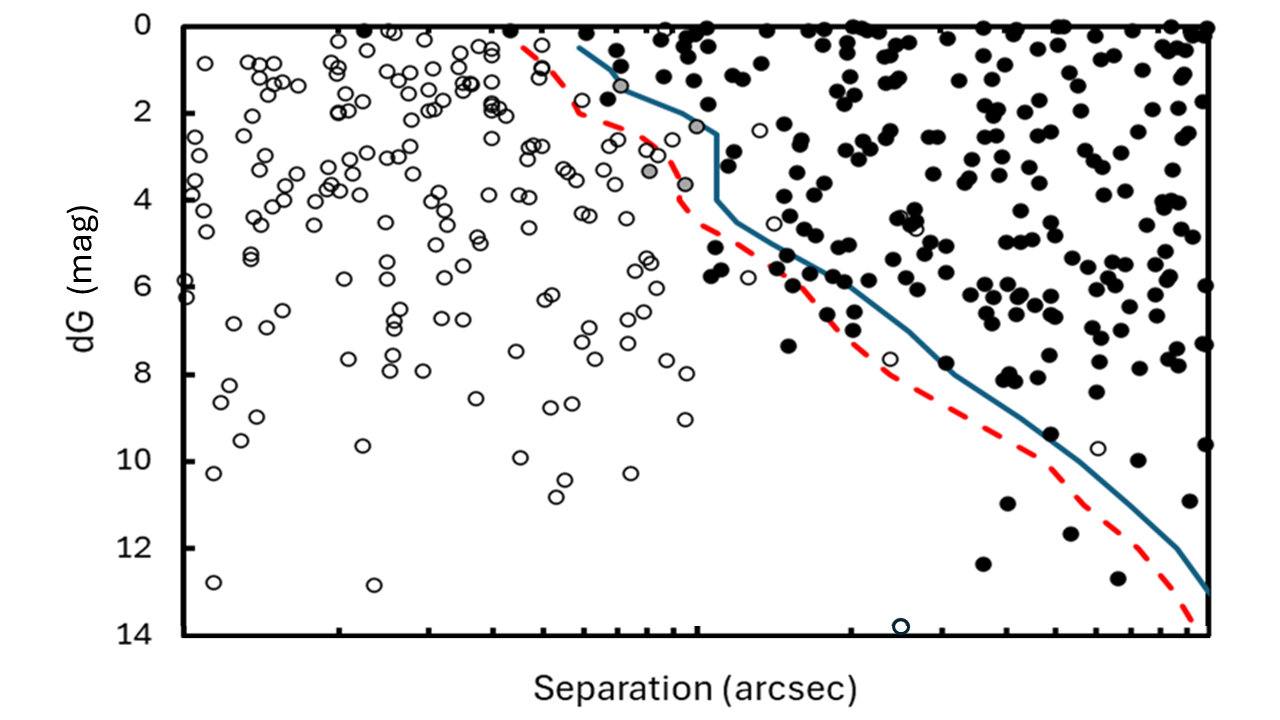}
\caption{Run of the contrast dG versus separation in arcsec for companions detected in Gaia DR3 in a sample of about 1300 young nearby star. We plotted only companions with separation $<10$ arcsec. Black filled circles are companions detected by Gaia DR3. Grey filled circles are companions for which Gaia DR3 gives only position, but not G magnitude. Empty circles are companions not detected in Gaia DR3. The blue solid and red dashed lines are the 50\% and 1\% contrast limits for Gaia DR2 according to \citet{Brandeker2019}, respectively }
\label{fig:gaia_dr3_limit}
\end{figure}

Figure \ref{fig:gaia_dr2_dr3} shows the run of the contrast $dG$ vs separation in arcsec for companions detected in Gaia DR3. We plotted only companions with separation $<10$ arcsec. Different symbols are for companions also detected with full astrometric solution also in Gaia DR2, those detected but without an astrometric solution in Gaia DR2, and finally for those detected in Gaia DR3 but not in Gaia DR2. Some companions close to the detection limits are detected with full astrometric solutions in Gaia DR3, while they are not detected or detected but without an astrometric solution in Gaia DR3. For comparison, we also show the 50\% and 1\% contrast limits for Gaia DR2 according to \citet{Brandeker2019}. The 1\% contrast limits for Gaia DR2 give a better representation of the actual limits for Gaia DR3. We attribute the higher detection rate of close companion in Gaia DR3 to the larger amount of data used in the most recent release.

To reinforce this argument, we also considered companions that were not detected by Gaia DR3. Furthermore, in order to have a wider statistics, we considered a sample of about 1300 nearby (distance $<100$ pc) young solar-type stars, for which we made a search for companions similar to that described in this paper. We found about 1100 companions around these stars. We then considered how the contrast $dG$ changes with apparent separation for these companions (see Figure \ref{fig:gaia_dr3_limit}). We used different symbols for those stars detected by Gaia DR3 and those that were not detected there. For the stars not detected by Gaia DR3, the $G$ magnitude of the companion was estimated using magnitudes at different wavelengths (e.g. from speckle interferometry or HCI) or, if these were not available, from masses (estimated using the same criteria described in Section \ref{sec:companion_parameters}) and the \citet{Baraffe2015} isochrones, with ages appropriate for each star. Whenever possible and relevant (periods in the range from years to hundreds of years), we considered the apparent separation at the epoch of the Gaia observation. If we now consider the Gaia DR2 1\% detection limit given by \citet{Brandeker2019}, we found that in this sample Gaia DR3 detected 10 companions fainter than this curve, while it did not detect nine stars brighter than this limit. Gaia DR3 detected 12 of the 16 stars that have companions brighter than the 1\% but fainter than the 50\% detection limit of Gaia DR2. 22 companions fainter than the 50\% detection limit of Gaia DR2 were detected by Gaia DR3, while only 5 brighter companions were not detected. This comparison supports our choice of using the 1\% Gaia DR2 detection limit by \citet{Brandeker2019} as a representative limit for the Gaia DR3 detections.

\section{Tables with data for the programme stars}
\label{appendix:data_tables}

This Appendix contains tables giving the relevant data about companions around the program stars. Table \ref{tab:photometry_ursa_major} contains the basic data. The first four columns contain the Hipparcos, HD, and 2MASS numbers as well as alternative names for each star. Column 5 gives the membership probability obtained using the Banyan-$\Sigma$ code \citep{Gagne2018}. Columns 6 and 7 give the coordinates. Column 8 the parallax. Whenever possible this was from Gaia DR3 \citet{Gaia_DR3}; else it was taken from Gaia DR2 or alternatively from the Hipparcos catalogue \citep{1997ESASP1200.....E}. Columns 9 and 10 give the Gaia $G$ magnitude of the primary $G_A$ and if existing of the secondary $G_B$, respectively. Column 11 gives the 2MASS $K$ magnitude \citep{2mass}. Columns 12 and 13 give the the absolute $M_G$ magnitude of the primary and secondary ($M_{GB}$). Column 14 gives the absolute $M_K$ magnitude of the primary. Column 15 gives the cluster, and Column 16 specifies if the star is in the core or tail of the cluster.

Table \ref{tab:info_ursa_major} explains if relevant information about binarity are available and Table \ref{tab:binary_ursa_major} gives the data relevant. The first four columns are as described above. Column 5 gives the Gaia $RUWE$ parameter; columns 6-10 give the Gaia mean RV, its error, the number of epochs considered in its derivation, the probability that variations around the mean value is due to random noise, and the robust RV amplitude. Columns 11 and 12 give the separation and position angle of visual companions. Column 13 gives the signal-to-noise ratio (SNR) of the Proper Motion anomaly (PMa) as given by \citet{Kervella2022}. Column 14 gives the minimum mass detectable using PMa.

Table \ref{tab:mass_ursa_major} gives the derived values for the stars and their companions. The first four columns are as described above. Column 5 and 6 gives the mass of the primary and of companions. Here, mass of the primary is the mass of all components within the semi-major axis of the companion considered. Column 7 gives the semi-major axis $a$. Column 8 gives the mass ratio $q$. Column 9 gives the detection method: either visual binary VIS if the secondary was detected with some imaging mode and a separate magnitude could be obtained; or dynamical binary DYN if the secondary is only detected from the dynamical impact on the primary. Finally, Column 10 gives references appropriate for each object, that are explained in the notes. Gaia means that the companion was detected as a separate entry in the Gaia catalogue; RUWE/RV/PMa and RUWE/RV means that the companion was detected from a high RUWE value ($>1.4$), a high scatter in RV ($>10$ \kms) or from a high SNR PMa ($>3$). All available data were used to derive a consistent solution, using the method described in \citet{Gratton2023}.

\begin{table*}
\caption{Photometry of stars. Only lines for the first five stars are shown; the full table is available at CDS.}
\scriptsize
\begin{tabular}{cccccccccccccccc}
\hline
HIP	&	HD	&	2MASS	&	Others	&	Prob	&	RA	&	Dec	&	$\pi$	&	$G_A$	&	$G_B$	&	$K_A$	&		$M_G$	&	$M_{GB}$	&	$M_K$ & Cluster & Group	\\
&&&&&degree&degree&mas&mag&mag&mag&mag&mag&mag\\
\hline
51814	&	91480	&	J10350969+5704578	&	37 UMa	&	99.9	&	158.7904	&	57.0826	&	37.34	&	5.062	&		&	4.334	&	2.923	&		&	2.195	& UMA & C \\
53910	&	95418	&	J11015046+5622566	&	$\beta$ UMa	&	99.8	&	165.4603	&	56.3824	&	38.60	&	2.399	&		&	2.350	&	0.332	&		&	0.283	& UMA & C \\
58001	&	103287	&	J11534983+5341408	&	$\gamma$ UMa	&	99.5	&	178.4577	&	53.6948	&	39.21	&	2.436	&		&	2.330	&	0.403	&		&	0.297	& UMA & C \\
59496	&	238087	&	J12120522+5855351	&	+59 1428	&	96.4	&	183.0218	&	58.9264	&	37.42	&	9.479	&		&	6.832	&	7.345	&		&	4.698	& UMA & C \\
59774	&	106591	&	J12152554+570157	&	$\delta$ Uma	&	99.9	&	183.8565	&	57.0326	&	40.51	&	3.311	&	13.421	&	3.090	&	1.349	&	11.459	&	1.128	& UMA & C \\
59774	&	106591	&	J12152554+570157	&	$\delta$ Uma	&		&		&		&		&		&	16.780	&		&		&	14.818	&		\\
\hline
\end{tabular}
\normalsize
\label{tab:photometry_ursa_major}
\end{table*}

\begin{table*}
\caption{Information about multiplicity. Only lines for the first five  stars are shown; the full table is available at CDS.}
\scriptsize
\begin{tabular}{lccccccccc}
\hline
HIP	&	HD	&	2MASS	&	Others		& Age&TESS &RV&HCI&RUWE&PMa\\
    &&&    &Myr &&&&&\\
\hline
51814	&	91480	&	J10350969+5704578	&	37 UMa	&	415	&	1	&	1	&	0	&	1	&	1	\\
53910	&	95418	&	J11015046+5622566	&	$\beta$ UMa	&	415	&	1	&	1	&	1	&	1	&	1	\\
58001	&	103287	&	J11534983+5341408	&	$\gamma$ UMa	&	415	&	1	&	0	&	1	&	0	&	0	\\
59496	&	238087	&	J12120522+5855351	&	+59 1428	&	415	&	1	&	2	&	0	&	1	&	1	\\
59774	&	106591	&	J12152554+570157	&	$\delta$ Uma	&	415	&	1	&	2	&	1	&	1	&	1	\\
\hline
\end{tabular}
\normalsize
\tablefoot{TESS=1 means there is TESS data, 0 there is not. RV=2 means there is high-precision RV data; RV=1 means there is Gaia RV data; RV=0 means there is no RV data. HCI=1 means there is HCI data, 0 there is not. RUWE=1 means there is RUWE data, 0 that there is not. PMa=1 means there is PMa data, 0 there is not.}
\label{tab:info_ursa_major}
\end{table*}

\begin{table*}
\caption{Binary data. Only lines for the first five  stars are shown; the full table is available at CDS.}
\scriptsize
\begin{tabular}{ccccccccccccc}
\hline
HIP	&	HD	&	2MASS	&	Others	&	RUWE	&	RV	&	RV err	&	Nb	&	RV ampl	&	sep	&	PA	&	SNR PMa	& $M_{min}$\\
&&&&& \kms & \kms && \kms & arcsec & degree & & $M_{\rm Jup}$\\
\hline
51814	&	91480	&	J10350969+5704578	&	37 UMa	&	0.998	&	-12.32	&	0.24	&	23	&	3.63	&		&		&	1.81	& 4.16\\
53910	&	95418	&	J11015046+5622566	&	$\beta$ UMa	&	5.991	&		&	0.03	&		&	0.19	&		&		&	2.20	&93.01\\
58001	&	103287	&	J11534983+5341408	&	$\gamma$ UMa	&		&		&		&		&		&	0.46	&		&		&\\
59496	&	238087	&	J12120522+5855351	&	+59 1428	&	1.936	&	-10.31	&	0.16	&	17	&	1.26	&		&		&	157.24	&1.69\\
59774	&	106591	&	J12152554+570157	&	$\delta$ Uma	&	2.635	&	-12.39	&	0.79	&	13	&	3.39	&		&		&	0.76	&20.80\\
59774	&	106591	&	J12152554+570157	&	$\delta$ Uma	&		&		&		&		&		&	590.18	&	39.49	&		\\
59774	&	106591	&	J12152554+570157	&	$\delta$ Uma	&		&		&		&		&		&	587.84	&	39.96	&		\\
\hline
\end{tabular}
\normalsize
\label{tab:binary_ursa_major}
\end{table*}

\begin{table*}
\caption{Mass and separation of companions. Only lines for the first five  stars are shown; the full table is available at CDS.}
\scriptsize
\begin{tabular}{cccccccccl}
\hline
HIP	&	HD	&	2MASS	&	Others	&	$M_A$	&	$M_B$	&	a	&	q	&	Method	&	Remarks	\\
	&		&		&		&	\MSun	&	\MSun	&	au	&		&		&		\\
\hline
51814	&	91480	&	J10350969+5704578	&	37 UMa	&	1.46	&		&		&		&		&		\\
53910	&	95418	&	J11015046+5622566	&	$\beta$ UMa	&	2.96	&		&		&		&		&		\\
58001	&	103287	&	J11534983+5341408	&	$\gamma$ UMa	&	2.82	&	1.355	&	11.73	&	0.480	&	VIS	&	$[6]$\\
59496	&	238087	&	J12120522+5855351	&	+59 1428	&	0.72	&	0.120	&	8.78	&	0.167	&	DYN	&	$[2]$	\\
59774	&	106591	&	J12152554+570157	&	$\delta$ Uma	&	2.13	&		&	0.36	&		&		&	$[1]$	\\
59774	&	106591	&	J12152554+570157	&	$\delta$ Uma	&	2.13	&	0.120	&	14568	&	0.056	&	VIS	&	$[1]$	\\
59774	&	106591	&	J12152554+570157	&	$\delta$ Uma	&	2.25	&	0.086	&	14511	&	0.038	&	VIS	&	$[1]$	\\
\hline
\end{tabular}
\normalsize
\tablefoot{[1]  Gaia DR3 \citep{Gaia_DR3}; [2]  RUVE-RV-PMa, this paper; [3]  \citet{Tokovinin2018}; [4]  \citet{Mason2001}; [5]  \citet{Pourbaix2004}; [6]  \citet{Stone2018}; [7]  \citet{Kiefer2018}; [8]  \citet{Grandjean2021}; [9]  \citet{Frankowski2009}; [10] \citet{Yelverton2019}; [11] \citet{Malkov2012}; [12] \citet{Talor2019}; [13] \citet{Damasso2023}; [14] \citet{Capistrant2024}; [15] \citet{Makarov2005}; [16] \citet{Trifonov2020}; [17] \citet{Schmitt2016}; [18] \citet{Galland2006}; [19] \citet{Lam2023}; [20] \citet{Quinn2016}; [21] \citet{Mermilliod2009}; [22] \citet{Demircan2006}; [23] \citet{Abt1999}; [24] \citet{Prsa2022}; [25] \citet{Newton2022}; [26] \citet{Guenther2005}; [27] \citet{Torres2024}; [28] \citet{Avvakumova2013}; [29] \citet{Patience1998}; [30] \citet{daSilva2006}; [31] \citet{Kaye1999}; [32] \citet{Bender2008}; [33] \citet{Tokovinin2014}; [34] \citet{Torres1997}; [35] \citet{Mason2009}; [36] \citet{Teng2023}; [37] \citet{Franson2023c}; [38] \citet{Abt1985}; [39] \citet{Diaz2012}; [40] \citet{Griffin1988}; [41] \citet{Debernardi2000};}

\label{tab:mass_ursa_major}
\end{table*}

\end{appendix}

\end{document}